\documentclass[%
 reprint, 
 amsmath,amssymb,
 aps, physrev,
]{revtex4-2}

\usepackage{graphicx}
\usepackage{dcolumn}
\usepackage{bm}

\begin{document}


\title{\textbf{Competition between acoustic radiation force and  streaming-induced drag force in focused beams for 3D cell trapping} 
}%

\author{Shiyu Li}
\affiliation{State Key Laboratory of Ocean Engineering, School of Ocean and Civil Engineering, Shanghai Jiao Tong University, Shanghai, 200240, China}


\author{Zhixiong Gong}
\email{Corresponding author: zhixiong.gong@sjtu.edu.cn}
\affiliation{State Key Laboratory of Ocean Engineering, School of Ocean and Civil Engineering, Shanghai Jiao Tong University, Shanghai, 200240, China}
\affiliation{Key Laboratory of Marine Intelligent Equipment and System, Ministry of Education, Shanghai, 200240, China}


\date{\today}

\begin{abstract}
The ability to trap a single cell or microparticle in three dimensions is important for biomedical and microfluidic applications. 
Single-beam acoustic tweezers based on focused waves offer a compact and biocompatible solution with their high spatial resolution and intensity gradient.
However, the three-dimensional (3D) trapping remains limited because the weak axial restoring radiation forces cannot always surpass the pushing drag force from the acoustic bulk streaming (ABS) particularly at high frequencies.
The combined effect of the acoustic radiation force (ARF, $F_{\rm rad}$) and the drag force induced by the ABS on a microparticle has not yet studied throughout in the free-space. It is known that the ARF depends on the square of the pressure amplitude at the focus as $p_{\text{foc}}^2$, whereas the drag force ($F_{\rm drag}$) induced by the ABS remains an open question with respect to $p_{\text{foc}}$ at different flow conditions.
In this study, we propose a unified theoretical–numerical framework to compare the contribution of the ARF and the drag force by the ABS, and systematacially derive an explicit scaling law for the streaming velocity $U_0 \sim p_{\text{foc}}^n$ with the hydrodynamic flow from the viscous to the inertial regimes.
It is found that the exponential power $n$ of $p_{\text{foc}}$ is bounded by $n=2$ for the viscous limit (the flow Reynolds number $Re_{\lambda} \ll 1$) and by $n=4/3$ for the inertial limit ($Re_{\lambda} \gg 1$), and $n \in (4/3,2)$ throughout the transition regime ($Re_{\lambda} \sim 1$). 
In addition, the Schiller–Naumann model is introduced to facilitate a more accurate estimation of the drag force compared to the Stokes model.
Based on the combined effect, we find that the trap ratio of the axial ARF over the drag force can vary nonmonotonically with $p_{\text{foc}}$, which contradicts the conventional expectation of the monotone increase. 
This work provides a theoretical foundation for the 3D trapping of a single cell with single-beam acoustical tweezers and a guidance for the optimized parameters. 
\end{abstract}

\maketitle


\section{\label{sec:introduction}Introduction}

Cells are the fundamental structural and functional units of living organisms. They regulate a wide range of biological processes essential for sustaining life. The precise 3D trapping of individual cell has therefore become a key tool in biomedical research, with applications including the measurement of cellular mechanical properties~\citep[]{zhou2025acoustic}, reproductive cell selection, and single-cell analysis~\citep[]{barrow2018natural,landsberg2012melanomas}.

Strategies for cell manipulation can be broadly classified into contact and non-contact methods. One of the contact-based methods called the micropipette aspiration technique employs the pressure differences to extract cells~\citep[][]{vles1933recherches,hochmuth2000micropipette}. However, these methods are often mechanically invasive and may increase the risk of chemical cross-infection~\citep[][]{yang2023acoustic}. 
In contrast, non-contact techniques use external physical fields, including optical, magnetic, and acoustic fields to manipulate cells without physical contact. Prominent examples include the optical tweezers~\citep[][]{ashkin1986observation}, the magnetic tweezers~\citep[][]{de2012recent}, and the acoustic tweezers~\citep[][]{baudoin2020acoustic}. Optical tweezers offer excellent spatial precision but may cause photothermal damage at high laser intensities~\citep[][]{blazquez2019optical,yang2023acoustic}. Magnetic tweezers typically require the attachment of magnetic labels, limiting their use in label-free applications~\citep[][]{neuman2008single}. Acoustic tweezers, by contrast, are label-free, biocompatible, and effective in optically opaque media.

Within acoustofluidics, single-beam acoustic tweezers based on focused beams generate highly localized fields that enable selective trapping of individual cell~\citep[][]{baudoin2020spatially,gong2022single,li2025reversing}. 
In contrast to standing wave configurations, which often require dual transducers or one transducer and one reflector, focused beams offer higher spatial selectivity and greater flexibility in experimental design and for the potential in vivo applications like the ultrasound imaging probe. 
Most human cells suspended in water exhibit a positive acoustic contrast factor and are thus trapped to the pressure minima~\citep[][]{gong2022single}. Hence, the acoustical focused vortices can be used because of the doughnut-shaped distributions of pressure amplitude with a central null. The trapping size of the ring-shaped potential well is on the order of the acoustic wavelength~\citep[][]{baresch2016observation,baudoin2020acoustic}. 
This facilitates the transducers to work at high frequencies for a small wavelength and high spatial resolution of a few micrometers.
Although the acoustic radiation force (ARF) can support the 3D trapping at such high frequencies~\citep[][]{gong2021three}, the  accompanying acoustic bulk streaming (ABS, also called the Eckart streaming)—the steady flow arising from viscous attenuation in the flow volume, exerts a hydrodynamic drag that may destabilise the trap. 
One commonly-used strategy is to place a cover glass above the focal region, which provides a supporting force against the drag force by the bulk streaming. However, this will limit the acoustic manipulation to a quasi-two-dimensional geometries with undesired physical contact ~\citep[][]{baudoin2020spatially}. 
Except for the focused vortex beams, the widely-used focused beam also show its potential for the 3D trapping of human cells.
This is implemented by introducing the iodixanol solution (one kind of biocompatible mediums) to reverse the acoustic contrast factor so that typical human cells can be trapped at the focus ~\citep[][]{li2025reversing}. Here the “contrast factor reversal” is used only as an intuitive description of the trapping tendency, the actual radiation force calculations in the present study are performed using the full angular spectrum scattering formulation, which naturally accounts for finite size effects beyond the Rayleigh limit.
However, likewise the case for the focused vortex beam, the strong bulk streaming can also push the trapped cell away and break the axial trap for focused beams, which can not be ignored and is not taken into consideration in previous works.

While the ARF is widely regarded as the principal mechanism for microparticle trapping, a unified treatment that incorporates drag force induced by the ABS has remained absent for single-beam acoustic tweezers based on focused beams. 
The acoustic streaming comes from the viscous dissipation either in boundary layers (for the Rayleigh streaming) or within the fluid bulk (for the ABS), and it induces a wave-following particle drift that can oppose trapping. ~\citep{li2024combined}. 
These flows are further amplified in high-frequency (exceeding tens of MHz) focused systems, where steep spatial gradients and strong energy localization intensify the streaming. 
As a result, streaming-induced drag force can match or even exceed ARF, particularly for very high frequencies. 
Existing studies have primarily examined the ARF–streaming interplay in microchannels driven by standing waves, where Rayleigh streaming dominates and the particle motion is constrained by channel walls \citep{nama2015numerical,muller2012acoustic}. 
However, these models do not apply to free-space focused beams, where (i) ABS arises from bulk absorption rather than boundary effects, (ii) particles are not geometrically confined and may escape axially, and (iii) the scaling of ABS with respect to focal pressure and transducer aperture remains unknown.
To our knowledge, a unified framework is still lacking that describes how ABS scales with acoustic pressure amplitude from viscous- to inertia-dominated regimes and that systematically contrasts the resulting pressure-dependent drag force with the strictly quadratic pressure dependence of the ARF. Furthermore, neglecting finite-particle Reynolds number corrections ($Re_s$), or relying solely on Stokes drag, can substantially underpredict the streaming-induced drag force in high-frequency focused tweezers.

In this work, we establish a unified theoretical–numerical framework that quantifies the combined effect between ARF ($F_{\rm rad}$) and drag force ($F_{\rm drag}$) induced by the ABS and assesses the feasibility of 3D trapping of representative human cells with the single-beam acoustic tweezers based on focused beams. The ARF is computed via the angular spectrum method, while AS is obtained from finite-element simulations of the steady Navier–Stokes equations with an acoustic body force source term. Scaling analyses further elucidate the crossover from viscous- to inertia-dominated streaming, revealing a non-monotonic dependence of the trap ratio $\Gamma = F_{\rm rad}/F_{\rm drag}$ on focal pressure. The resulting framework provides predictive capability and new physical insight into the competition between ARF and ABS-induced drag force in free-space focused beams, offering practical design guidelines for achieving stable 3D trapping of a single cell with single-beam acoustic tweezers.

The remainder of the paper is organized as follows. In Section~\ref{sec:Problem formulation and governing equations}, we describe the systematic configuration and governing equations, and introduce the two primary nonlinear mechanisms of interest: the acoustic radiation force and the acoustic bulk streaming. 
Section~\ref{sec:numerical methods} outlines the numerical approach, comprising the angular spectrum method for computing the acoustic radiation force and a finite element method for solving the streaming field based on our derived source term ~\citep[]{li2025reversing}. 
The main results are presented in Section~\ref{sec:results and discussion}. 
 Section~\ref{sec:Flow characteristics} examines the flow characteristics of the bulk streaming field, Section~\ref{sec:transducer} explores the influence of transducer geometry on both the radiation force and the streaming-induced drag force, and Section~\ref{sec:pressure} quantifies the relative magnitude of acoustic radiation force and drag force across a range of focal pressures. 
 A summary of key findings and their implications for device design is provided in Section~\ref{sec:conclusions}. 
 In this study, we consider linear acoustic propagation in homogeneous Newtonian fluids with small Mach numbers,  neglecting thermal effects and material inhomogeneities.

\section{Problem formulation and governing equations}\label{sec:Problem formulation and governing equations}
\begin{figure}
\centering
\includegraphics[width=0.5\textwidth]{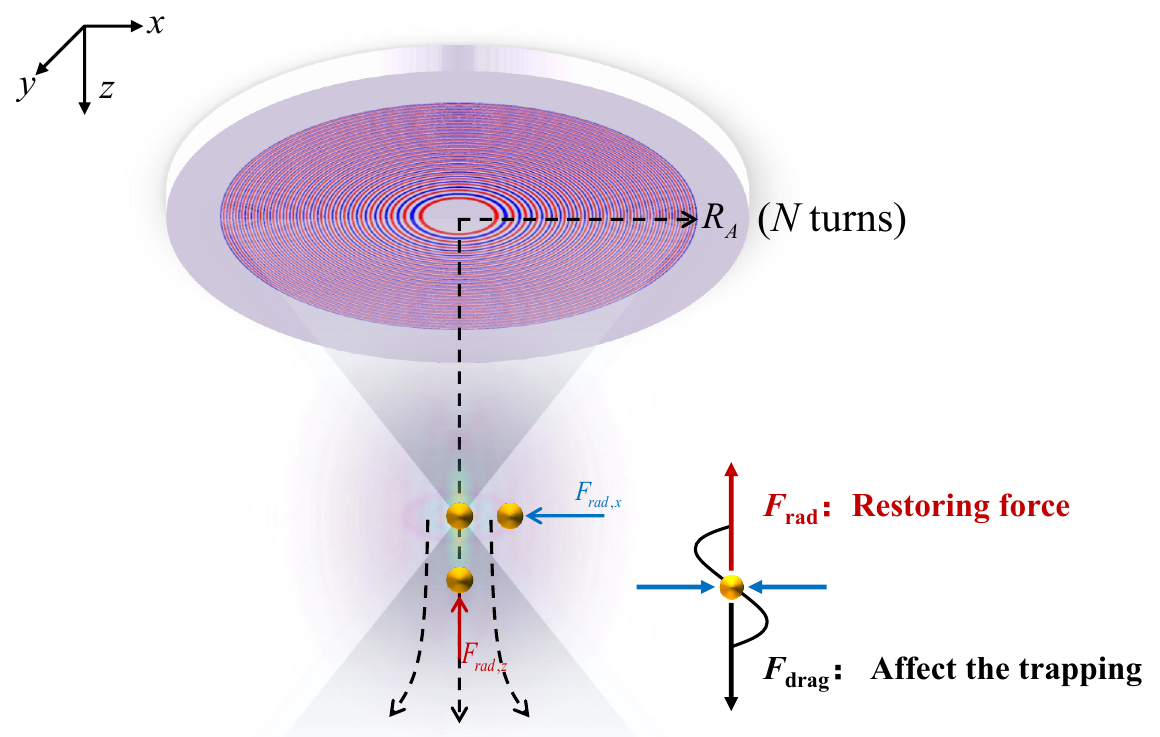}
\caption{
Schematic of the focused beam. Two sets of electrodes (red and blue) on a piezoelectric wafer are excited with antiphase (i.e., with a phase difference of $\pi$). The aperture radius $R_A$ is determined by the number of electrode turns $N$, with representative parameters given in reference~\citep[]{li2025reversing}. The acoustic energy distribution is shown as a gradient field, with energy strongly localized near the focal point. Restoring forces along the axial and lateral directions are indicated by red and blue arrows, respectively, enabling 3D trapping of cells (depicted as golden spheres). Suspended cells in the acoustic field experience a combination of radiation force (red arrow) and drag force arising from acoustic streaming (black arrow).}
\label{Fig: Sketch}
\end{figure}

The configuration of the focused beam system is illustrated in~Fig.\ref{Fig: Sketch}. The planar focused ultrasound transducer employed in this study operates at a center frequency of $f = 40$~MHz and produces focused acoustic beams with a focal length of $h = 1$~mm. The aperture radius $R_A$ is determined by the number of turns $N$ and the focal length $h$ in the transducer, for instance, $N=26$ with $h$=1~mm gives $R_A$=1.72~mm. In the present MEMS design, the turn number $N$ determines the effective aperture radius $R_A$ and therefore indirectly controls the focal structure of the generated beam.
The detailed specifications of the transducer can be found in reference~\citep[]{li2025reversing}. The acoustic tweezers were fabricated via standard micro-electro-mechanical systems (MEMS) techniques~\citep[][]{baudoin2019folding, baudoin2020spatially}, enabling a planar geometry compatible with microfluidic platforms and channels, and thus well suited for microscale manipulation.

We consider a homogeneous, compressible Newtonian fluid governed by the conservation of mass and momentum. The corresponding compressible Navier–Stokes equations are~\cite{bruus2012acoustofluidics}:
\begin{equation}
\frac{\partial \rho}{\partial t} + \nabla\cdot(\rho \boldsymbol{v}) = 0,
\label{Eq. 1: mass conservation}
\end{equation}
\begin{equation}
\frac{\partial (\rho \boldsymbol{v})}{\partial t} + \nabla\cdot (\rho \boldsymbol{v} \otimes \boldsymbol{v}) = -\nabla p + \mu_s \nabla^2 \boldsymbol{v} + \left( \frac{\mu_s}{3} + \mu_b \right) \nabla (\nabla\cdot \boldsymbol{v}),
\label{Eq.2: momentum conservation}
\end{equation}

Here, $\rho$, $p$, and $\boldsymbol{v}$ represent the fluid density, pressure, and velocity fields, respectively, while $\mu_s$ and $\mu_b$ are the dynamic viscosity and bulk viscosity. The system is closed by the following state equation:
\begin{equation}
\begin{aligned}
&p=p(\rho), \text { with }\left.\frac{\partial p}{\partial \rho}\right|_{s}=c_{0}^{2} \label{Eq.3: state equation}
\end{aligned}
\end{equation}

Owing to the nonlinearity of the governing equations, we employ a regular perturbation expansion in a small nondimensional parameter $\varepsilon$, taken to represent the acoustic Mach number~\citep[]{friend2011microscale}. The physical fields are expanded as:
\begin{equation}
\rho = \rho_0 + \varepsilon \rho_1 + \varepsilon^2 \rho_2, \label{Eq.3: rho}
\end{equation}
\begin{equation}
p = p_0 + \varepsilon p_1 + \varepsilon^2 p_2, \label{Eq.4: p}
\end{equation}
\begin{equation}
\boldsymbol{v} = \varepsilon \boldsymbol{v}_1 + \varepsilon^2 \boldsymbol{v}_2, \label{Eq.5: v}
\end{equation}
where $\rho_0$ and $p_0$ denote the ambient (equilibrium) density and pressure. The subscripts 1 and 2 correspond to first- and second-order perturbation fields, respectively. Assuming an initially quiescent fluid, the zeroth-order velocity vanishes.

Within this perturbative framework, the governing equations are separated into first-order and second-order systems, allowing for decoupled analysis of the oscillatory acoustic field and steady streaming flow. The linearized first-order equations for the acoustic perturbations 
$(\rho_1,\boldsymbol v_1,p_1)$ are~\cite{bruus2012acoustofluidics}:
\begin{equation} \begin{gathered} \frac{\partial \rho_1}{\partial t} + \rho_0 \nabla \cdot \boldsymbol{v}_1 = 0, \\ \rho_0 \frac{\partial \boldsymbol{v}_1}{\partial t} = -\nabla p_1 + \mu_s \nabla^2 \boldsymbol{v}_1 + \left( \frac{\mu_s}{3} + \mu_b \right) \nabla (\nabla \cdot \boldsymbol{v}_1), \end{gathered} \end{equation}
where $\nabla^2$ denotes the Laplace operator. The first-order state equation follows from a Taylor expansion: $p_1 = c_0^2 \rho_1$.
Due to the harmonic nature of the fields, the time-domain equation above can also be expressed in the following frequency-domain form:
\begin{subequations}
\begin{align}
    -i \omega \rho_1 &= -\rho_0 \nabla \cdot \boldsymbol{v_1} \label{Eq:8a} \\
    -i \omega \rho_0 \boldsymbol{v_1} &= -\nabla p_1 
    + \mu_s \nabla^2 \boldsymbol{v}_1 
    + \left( \frac{\mu_s}{3} + \mu_b \right) \nabla (\nabla \cdot \boldsymbol{v}_1) \label{Eq:8b}
\end{align}
\end{subequations}
Taking the divergence on both sides of Eq.\eqref{Eq:8b}, substituting into Eq.\eqref{Eq:8a}, and incorporating the state equation yields the following Helmholtz equation in viscous fluid:
\begin{equation}
\label{Eq:helmholtz}
\nabla^2 p_1+k^2 p_1=0
\end{equation}
The complex wavenumber $k$ is defined by the following expression:
\begin{equation}
k=\frac{\omega}{c_0}\left[1-\frac{i \omega}{\rho_0 c_0^2}\left(\frac{4}{3} \mu_s+\mu_b\right)\right]^{-\frac{1}{2}}
\end{equation}

The second-order steady continuity equation reads:

\begin{equation}
\rho_0 \nabla \cdot \boldsymbol{v}_2 = -\nabla \cdot (\rho_1 \boldsymbol{v}_1),
\end{equation}
We neglect the right-hand side, which accounts for weak compressibility effects at small Mach number, and obtain an incompressible streaming flow, $\nabla \cdot \boldsymbol{v}_2 = 0$ (Nyborg’s approximation~\cite{nyborg1958acoustic,nyborg1965acoustic}).

The second-order momentum equation becomes:
\begin{equation}
\begin{split}
& \rho_{1} \frac{\partial \mathbf{v_{1}}}{\partial t}+\rho_{0} \mathbf{v_{1}} \cdot \nabla \mathbf{v_{1}} \\
& =-\nabla p_{2}+\mu_s \Delta \mathbf{v_{2}} + \left (\frac{\mu_s}{3}+\mu_b \right) \nabla \nabla \cdot \mathbf{v_{2}}
\label{Eq. A7: second order Momentum Conservation}
\end{split}
\end{equation}

After simplification, the steady second-order momentum equation reduces to:
\begin{equation}
\label{Eq:streaming equation}
-\nabla p^*_2+\mu_s \Delta \mathbf{v}_2 + \mathbf{F_s} = 0
\end{equation}
where $\mathbf{F_s}  = \left({4 \mu_s}/{3}+\mu_b \right) {\omega^2}\left\langle p_1 \mathbf{v}_1\right\rangle/({c_0^4 \rho_0})$ is the sole source of the acoustic streaming. $\nabla p^*_2 =\nabla  p_2 + \nabla\langle \mathcal{L} \rangle$, and $\langle \mathcal{L} \rangle$ is the average acoustic Lagrangian. The details of these equations can be found in Ref.\cite{baudoin2020acoustic}.

The total force $\mathbf{F_p}$ acting on a suspended particle is evaluated by integrating the stress tensor $\boldsymbol{\sigma}$ over the particle surface $S_{\mathrm{p}}(t)$:
\begin{equation}
{\boldsymbol{F}}_{\mathbf{p}} = \left\langle \iint_{S_{\mathrm{p}}(t)} \boldsymbol{\sigma} \cdot \boldsymbol{n}_{\mathbf{p}} \, \mathrm{d}S \right\rangle,
\end{equation}
where $\boldsymbol{n}_{\mathbf{p}}$ denotes the outward unit normal to the particle surface, and $S_{\mathrm{p}}(t)$ denotes the surface of the oscillating particle. The stress tensor $\boldsymbol{\sigma}$ is defined as:
\[
\boldsymbol{\sigma} = -p \boldsymbol{I} + 2\mu_s \boldsymbol{D} + \left( \mu_b - \tfrac{2}{3}\mu_s \right)(\nabla \cdot \boldsymbol{v}) \boldsymbol{I}.
\]
where $\boldsymbol{D}$ is the strain tensor, and $\boldsymbol{I}$ is the identity tensor.

This force ${\boldsymbol{F}}_{\mathbf{p}}$ can be decomposed into two components:
\begin{equation}
{\boldsymbol{F}}_{\mathbf{p}} ={\boldsymbol{F}}_{\mathbf{rad}} + {\boldsymbol{F}}_{\mathbf{str}},
\end{equation}
where ${\boldsymbol{F}}_{\mathbf{rad}}$ is the acoustic radiation force:
\begin{equation}
{\boldsymbol{F}}_{\mathbf{rad}} = \iint_{S_{\mathrm{R}}} \left( -\rho_0 \langle {\boldsymbol{v}_1} \otimes {\boldsymbol{v}_1} \rangle + \langle \mathcal{L} \rangle \right) \cdot \boldsymbol{n}_{\mathbf{R}} \, \mathrm{d}S,
\label{Eq:Frad general formulation}
\end{equation}
Here, $\langle \mathcal{L} \rangle$ is the time-averaged acoustic Lagrangian \citep[][]{baudoin2020acoustic, li2024eckart} and $S_R$ denotes a closed surface at rest enclosing the particle. Under plane wave incidence, the magnitude of the acoustic radiation force $F_{\rm rad}$~follows the classical quadratic dependence on pressure amplitude~\cite{hasegawa2001frequency}. Taking the focal
pressure \(p_{\mathrm{foc}}\) as the characteristic amplitude gives
\begin{equation}
{F}_{\rm rad} \;\propto\; p_{\mathrm{foc}}^{\,2}.
\label{Eq:Frad scaling}
\end{equation}
The proportionality constant depends on material contrast, the dimensionless frequency \(ka\), and absorption, but the quadratic dependence on pressure amplitude remains.

The second term, ${\boldsymbol{F}}_{\mathbf{str}}$ represents the viscous drag exerted by the streaming flow and is given by:
\begin{equation}
{\boldsymbol{F}}_{\mathbf{str}} = \iint_{S_{\mathrm{R}}} \left( -p^*_2 \boldsymbol{I} + 2\mu_s \boldsymbol{D} \right) \cdot \boldsymbol{n}_{\mathbf{R}} \, \mathrm{d}S.
\end{equation}
This contribution is conventionally quantified through the classical hydrodynamic drag formulation in fluid mechanics, denoted as ${\boldsymbol{F}}_{\mathbf{drag}}$, which for a spherical particle in viscous flow takes the form of the Stokes drag force (see Eq.\eqref{Eq:Stokes Fdrag}).

It is therefore important to note that particles suspended in a fluid medium experience two conceptually distinct contributions in the present model: (i) the leading-order acoustic radiation force ${\boldsymbol{F}}_{\mathbf{rad}}$ arising from momentum transfer between the incident acoustic field and the particle, and (ii) the viscous drag force  ${\boldsymbol{F}}_{\mathbf{drag}}$, generated by acoustic streaming. The purpose of the present study is to isolate the competition between these two dominant effects in focused beam bulk trapping. A full thermoviscous treatment of the radiation-force term is beyond the scope of the present work and will be addressed in future studies.

The theoretical analysis herein is based on the following assumptions:
(i) A small acoustic Mach number, ensuring the validity of first- and second-order perturbation theory;
(ii) A Newtonian fluid with constant properties; 
(iii) Nyborg's approximation (\( \nabla\!\cdot\! \boldsymbol{v}_2=0 \));
(iv) Negligible thermo-viscous coupling and temperature-driven body forces; and
(v) Single-frequency excitation without significant higher-harmonic content in the focal region.
(vi) The radiation force is evaluated using a leading-order inviscid scattering model, and thermoviscous corrections to ${\boldsymbol{F}}_{\mathbf{rad}}$ are not included explicitly.

 Building on the present research~\citep{qiu2021fast,daru2021acoustically,das2025acoustothermal,zhang2011geometrical,marston2013viscous,marston2016unphysical,marston2017relationship,winckelmann2023acoustic}, we will progressively relax the current assumptions to develop a more comprehensive acousto-thermal-fluidic multiphysics model for focused beams. In particular, future work will incorporate thermoviscous boundary layer corrections to the radiation force term, particle surface dissipation, internal absorption, and near particle microstreaming, so as to assess their quantitative influence on trapping stability beyond the leading-order model adopted here.

\section{Numerical methods: ASM and FEM}\label{sec:numerical methods}

The acoustic radiation force is computed using a custom MATLAB implementation of the angular spectrum method (ASM) \citep[see][]{gong2022single}, whereas the streaming flow is simulated via finite element method (FEM) computations in COMSOL Multiphysics 6.2~\citep[][]{bach2018theory,muller2012numerical}. The acoustic tweezer system is configured to operate at a center frequency of $f=40\,\mathrm{MHz}$, designed to trap MCF-7 human breast cancer cells. For computational efficiency, the cells are modelled as homogeneous fluid spheres, enabling efficient evaluation of scattering coefficients~\citep[][]{li2025reversing}.

The suspending medium is a 60\% (v/v) iodixanol solution, selected to invert the acoustic contrast factor $\Phi_{\mathrm{SW}}$ of the cells. The fluid properties, including density, sound speed, and viscosity, are derived from empirical fits to experimental data~\citep[][]{augustsson2016iso}. The dynamic viscosity $\mu_s$ at this concentration is extrapolated from the fitted curve, while the bulk viscosity $\mu_b$ is assumed to be that of pure water due to the lack of direct measurement data \citep[][]{karlsen2016acoustic}. The physical parameters used in this study are summarized in Table~\ref{tab:Physical parameters}.

Unless otherwise stated, all pressure amplitudes are reported at the focal point and normalised to \(p_{\rm foc}=1\,\mathrm{MPa}\) for controlled comparisons.
The normalisation procedure are as follows: 
we first assign a source plane pressure magnitude \(p_{\rm src}^{(0)}=1\,\mathrm{MPa}\) and compute the acoustic field using the ASM. 
The resulting focal pressure magnitude \(p_{\rm foc}^{(0)}\) is then extracted, and the source pressure magnitude $p_{\rm src}$ is rescaled as
\begin{equation}
  p_{\rm src} = p_{\rm src}^{(0)}\,\frac{p_{\rm foc}}{p_{\rm foc}^{(0)}} .
\end{equation}
The field is subsequently recomputed such that the final focal pressure magnitude satisfies $p_{\rm foc}$=1~MPa.

In the following subsections, the two numerical methodologies are described: the angular spectrum method for evaluating the acoustic radiation force and the finite element method for modelling acoustic streaming.
\begin{figure} 
\centering 
\includegraphics[width=7.6cm]{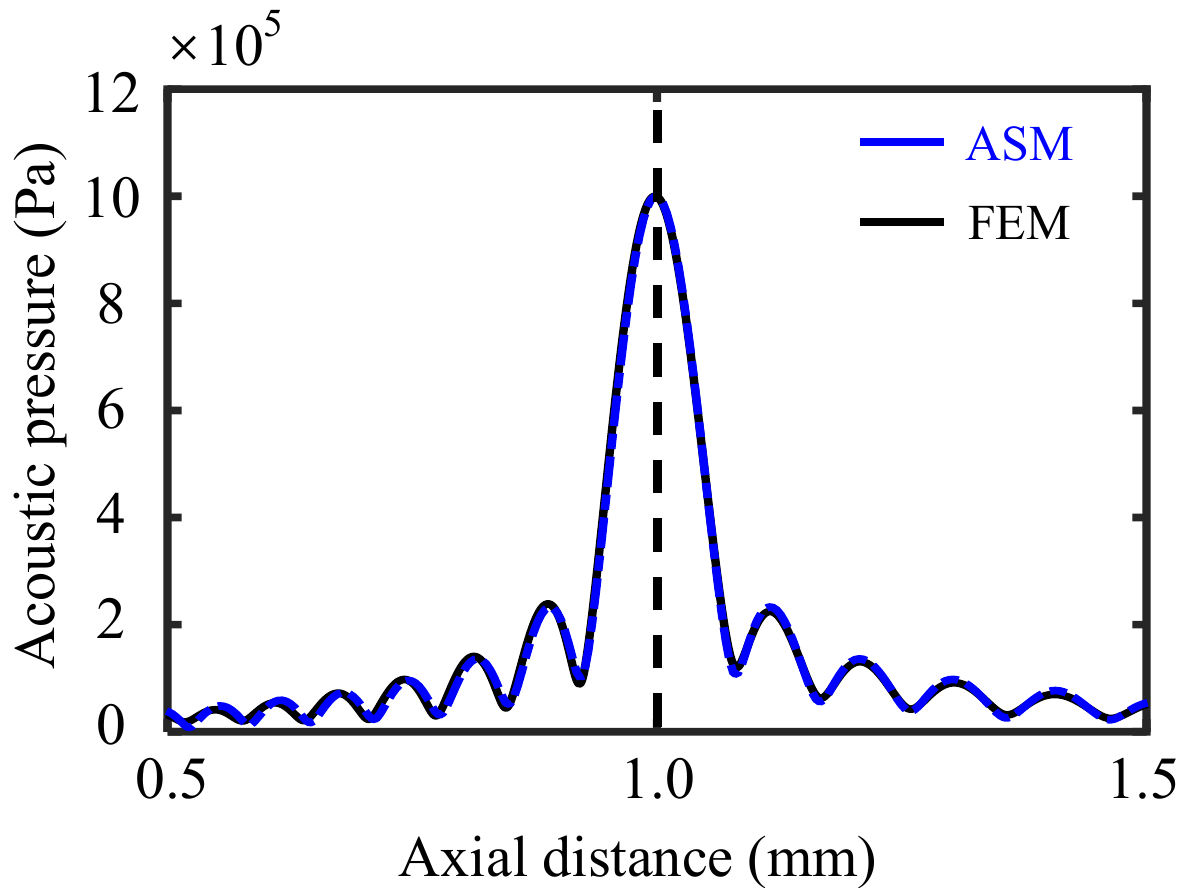}
\caption{
Axial distribution of the acoustic pressure magnitude predicted by FEM and ASM for a focused beam at 40 MHz, The focal pressure is normalized to $p_{\rm foc}$=1~MPa, and the focal length is $h$=1~mm. The dashed line marks the focal position. The two methods exhibit close agreement under identical conditions, thereby validating consistency between the numerical approaches.
}
\label{Fig: ASM & FEM}
\end{figure}
\subsection{Acoustic radiation force via the angular spectrum method}

The angular spectrum method represents arbitrary acoustic beams as a superposition of plane waves with different propagation directions. By incorporating plane wave scattering theory, ASM enables efficient evaluation of acoustic radiation forces on particles of arbitrary size and shape, extending beyond the Rayleigh limit. The radiation force on a spherical particle in a general acoustic field is expressed following Sapozhnikov and Bailey~\citep[]{sapozhnikov2013radiation}:

\begin{subequations}\label{ASM_force}
\begin{align}
F_x &=
\frac{1}{4\pi^{2}\rho_{0}k^{2}c_0^{2}}
\Re\Biggl\{
\sum_{n=0}^{\infty}\sum_{m=-n}^{n}
C_n \Bigl[
-\, b_{n+1}^{-m} H_{nm} H_{n+1,m-1}^{*} \notag\\
&\qquad\qquad\qquad\qquad
+\, b_{n+1}^{m}   H_{nm} H_{n+1,m+1}^{*}
\Bigr]
\Biggr\}, \label{ASM_Fx} \\[0.4em]
F_y &=
\frac{1}{4\pi^{2}\rho_{0}k^{2}c_0^{2}}
\Im\Biggl\{
\sum_{n=0}^{\infty}\sum_{m=-n}^{n}
C_n b_{n+1}^{m} \Bigl[
H_{n,-m} H_{n+1,-m-1}^{*} \notag\\
&\qquad\qquad\qquad\qquad
+\, H_{nm}   H_{n+1,m+1}^{*}
\Bigr]
\Biggr\}, \label{ASM_Fy} \\[0.4em]
F_z &=
-\frac{1}{2\pi^{2}\rho_{0}k^{2}c_0^{2}}
\Re\Biggl\{
\sum_{n=0}^{\infty}\sum_{m=-n}^{n}
C_n c_{n+1}^{m} H_{nm} H_{n+1,m}^{*}
\Biggr\}. \label{ASM_Fz}
\end{align}
\label{Eq:ASM Frad}
\end{subequations}

Here, \( H_{nm} \) denotes the beam-shape coefficients determined by the incident wave, \( C_n \) is a coefficient related to the dimensionless partial wave scattering coefficient, with \( b_n^m \) and \( c_n^m \) being the corresponding expansion coefficients associated with the mode indices \( n \) and \( m \). In the long-wavelength limit, the formulation in~Eq.\eqref{ASM_force} reduces to the classical Gor’kov theory. The procedure for computing the acoustic radiation force on particles of arbitrary size based on the ASM is detailed by Li and Gong (2025). Throughout this study, the magnitude of the radiation force is reported with respect to the minimum axial restoring force. In general, the lateral radiation force component exceeds the axial contribution, which provides stronger trapping capability.

In the present model, the expression Eq.\eqref{ASM_force} should be understood as a leading-order radiation force model, only the bulk viscous attenuation of the host fluid on wave propagation is included, whereas near particle thermoviscous boundary layer corrections to the scattering coefficients entering $F_{\rm rad}$ are not explicitly considered. Under the present parameter conditions, the viscous Stokes boundary layer thickness $\delta$ remains much smaller than the particle radius $a$ ($\delta \ll a$) in the Mie regime, i.e. the thin boundary layer limit, so these thermoviscous corrections are treated here as higher-order relative to the leading-order scattering contribution.
The focal pressure magnitude is taken as the common reference for both the radiation-force evaluation and the Eckart streaming-induced drag estimation, so that the comparison between the two effects is made under the same focal-field amplitude condition.
\subsection{Acoustic streaming via finite element modeling}

To exploit the time-scale separation between the rapidly oscillating acoustic field and the slowly evolving streaming flow, a two-step computational strategy is employed. In the first step, the acoustic pressure field is computed, and in the second, the resulting body force are used as source terms to drive the time-averaged acoustic streaming~\citep[]{muller2012numerical}.
Since accurately resolving high-frequency acoustic waves requires a minimum of six mesh elements per wavelength ~\citep[]{mace2008modelling}, making fully 3D simulations prohibitively expensive at frequencies of tens of megahertz. By exploiting the axial symmetry of the focused beam and balancing computational efficiency with accuracy, a two-dimensional axisymmetric finite element model is formulated for focused beam configurations. 

Simulations were carried out in \textsc{COMSOL Multiphysics} using a coupled multiphysics framework. The acoustic field was modelled with the "Pressure Acoustics, Frequency Domain" interface (governed by Eq.\eqref{Eq:helmholtz}) with Perfectly Matched Layers (PMLs) to represent an unbounded domain. The streaming flow was obtained from the "Laminar Flow" interface (governed by Eq.\eqref{Eq:streaming equation}) under open boundary conditions, to ensure non-reflecting outflow in a semi-infinite geometry.  The finite element results were verified through a mesh convergence study. 
For this purpose, a representative configuration with a transducer of 
$N=26$ turns and a focal length of $h=1~\mathrm{mm}$ was selected for the grid-refinement test. Convergence was then evaluated on a systematically refined sequence of meshes and reported using a successive-difference criterion.
Let $U_0^{(k)}$ denote the peak streaming velocity on grid $k$, and let $k+1$ represent the next finer grid, the relative difference is then defined as
\begin{equation}
   \Delta_{k \to k+1} =
   \frac{\bigl| U_0^{(k)} - U_0^{(k+1)} \bigr|}
        {\bigl| U_0^{(k+1)} \bigr|}
   \times 100\% ,
\end{equation}
which provides a straightforward measure of grid independence. 
The mesh resolution is expressed in terms of cells per wavelength (CPW), defined as the number of finite elements per acoustic wavelength $\lambda$, with \ $\mathrm{CPW}=\lambda/h_{mesh}$ where $h_{mesh}$ is the characteristic element size. In our refinement sequence 
($\mathrm{CPW}=8$--16), the relative difference between the two finest grids is 
$\Delta \leq 0.2\%$ (Table~\ref{tab:Mesh}), 
confirming that the predictions are mesh independent. In the subsequent simulations, we adopted the finest grid level 5 (L5), 
with a fluid domain of radial $\times$ axial extent 
$4~\mathrm{mm}\times 3~\mathrm{mm}$ and a perfectly matched layer (PML) 
of thickness equal to one wavelength.

\begin{table}
   \begin{center}
\label{tab:Mesh}
  \caption{Mesh convergence study for the streaming simulation. (Relative change $\Delta$ with respect to the next finer grid).}
  \begin{tabular*}{0.48\textwidth}{@{\extracolsep{\fill}}lll}
    \hline 
        \hline 
     Level  & CPW     & \(\Delta\) to next finer (\%) \\
    \hline 
     L1 & 8 &     \\
    L2 & 10  & 0.49  \\
     L3 & 12   & 0.34   \\
    L4 & 14   & 0.22   \\
    L5 & 16   & 0.18   \\
    \hline
        \hline 
  \end{tabular*}
   \end{center}
\end{table}

Once the steady streaming velocity field is obtained, the hydrodynamic drag force on a spherical particle is evaluated using the Schiller–Naumann (S–N) model, which remains valid for particle Reynolds numbers $Re_{s}$ up to $\textit{O}(10^{3})$~\citep[]{schiller1933uber}:
\begin{equation}
F_{\mathrm{drag}} = 6 \pi \mu_s R v_r \left( 1 + 0.15\, Re_s^{0.687} \right),
\label{Eq:Fdrag}
\end{equation}
Here, $R$ denotes the particle radius, $\mu_s$ is the dynamic viscosity of the medium, and $v_r$ is the magnitude of the relative velocity between the streaming flow ${\boldsymbol{v}}_{\mathrm{2}}$ and the particle ${\boldsymbol{v}}_{\mathrm{p}}$. The particle Reynolds number $Re_s$ is defined based on the particle diameter $d = 2R$ as:
\begin{equation}
Re_s = \frac{2 R \rho_m v_r}{\mu_s},
\end{equation}
where $\rho_m$ is the density of the suspending medium and $v_r$ is the relative velocity magnitude. The S–N drag model is particularly relevant to Mie particles in laminar flow conditions. For $Re_s \ll 1$, the S-N model reduces to the classical Stokes expression ~\citep[]{stokes1851effect}:
\begin{equation}
   F_{\mathrm{drag}} = 6\pi \mu_s R v_r. 
   \label{Eq:Stokes Fdrag}
\end{equation}

Under high acoustic pressures that produce strong streaming flows, the finite-$Re_s$
corrections from the Schiller–Naumann model become significant~\cite{cicek2017acoustophoretic}. This correction leads to a nonlinear relationship between drag force and velocity, consequently causing the drag force to exhibit complex variation with acoustic pressure~\citep[]{li2019axial}, the detailed analysis is presented in section~\ref{sec:pressure}.
\subsection{Combined force analysis for trapping evaluation}
\label{sec:combined_force_analysis}

To assess the feasibility of 3D particle trapping in high-frequency focused acoustic fields, it is essential to consider the combined effect between two dominant forces. The acoustic radiation force, ${\boldsymbol{F}}_{\mathbf{rad}}$, drives particles towards pressure nodes or antinodes depending on their contrast factor, and acts as a restoring force that promotes confinement. Conversely, the steady drag force induced by acoustic streaming, ${\boldsymbol{F}}_{\mathbf{drag}}$ tends to transport particles along the mean flow direction, thereby opposing stable trapping. The net balance between these competing forces ultimately determines the achievable trapping stability.

Before comparing the acoustic radiation force and drag force, it is essential to verify the consistency of the numerical methods used to compute the acoustic field, as both forces ultimately depend on the accuracy of the acoustic pressure field, ensuring consistency is a prerequisite for meaningful comparison. As shown in~Fig.\ref{Fig: ASM & FEM}, the acoustic pressure distributions obtained from the ASM and the FEM exhibit excellent agreement under identical transducer configurations. This confirms the validity of the hybrid modelling strategy, whereby the radiation force is computed from the ASM, while the streaming flow and corresponding drag force are evaluated using FEM. The demonstrated agreement ensures that subsequent force comparisons are grounded in a coherent and physically consistent framework.

To quantify the relative strength of the competing effects, we define the dimensionless trapping ratio $\Gamma$
\begin{equation}
\Gamma = \frac{{{F}}_{\mathrm{rad}}}{{{F}}_{\mathrm{drag}}},
\label{Eq:Gamma}
\end{equation}
where $\Gamma >1$, the radiation force dominates over the streaming drag and enables stable trapping at the focus. Conversely, when $\Gamma <1$, the drag force prevails, displacing the particle from the focal region.

We begin by systematically evaluating the influence of transducer parameters on both the acoustic radiation force and the streaming drag, from which an optimised configuration is identified. For this selected geometry, we then examine how the focal pressure governs
$\Gamma$, thereby informing the design of high-frequency acoustic tweezers. Importantly, the two forces exhibit distinct pressure scalings: the acoustic radiation force increases quadratically with pressure amplitude, while the streaming-induced drag force follows a more intricate, nonlinear dependence. The pressure scaling in the viscous and inertial limits is known~\citep{moudjed2014scaling,slama2019characterization}, but the transition regime, where focused beams most often operate, has lacked quantitative characterization. This mismatch in scaling exponents allows pressure tuning to shift the force balance towards radiation force dominated regimes. A detailed analysis of this scaling behaviour and its implications for trap design is presented in section~\ref{sec:pressure}.

\begin{table}
  \begin{center}
    \caption{Physical parameters.}
  \begin{tabular}{lccc}
   \hline 
       \hline 
      Parameter  & Symbol   &   Value  \\[3pt]
      \hline 
      Frequency & $f$ & \(40\,\mathrm{MHz}\) \\
       Wavenumber & $k$ & $2\pi f/c_{m}$ \\
       Wavelength & $\lambda$ & $c_{m}/f$ \\
       Attenuation coefficient & $\alpha$ & $\omega^2 \mu_s b/2 \rho_0 c_{0}^{3} $ \\
       Defined coefficients & $b$ &  $4/3 + \mu_b/\mu_s$ \\
       Fluid density   &$\rho_{m}$ & ~~\(1319.7\,\mathrm{kg\,m^{-3}}\)~ \\
       Sound speed of fluid   & $c_{m}$ & ~~\(1497.7\,\mathrm{m\,s^{-1}}\) \\
       Cell density   &$\rho_{c}$ & ~~\(1068\,\mathrm{kg\,m^{-3}}\)~ \\
       Sound speed of cell   & $c_{c}$ & ~~\(1497.7\,\mathrm{m\,s^{-1}}\) \\
       Dynamic viscosity  & $\mu_{s}$ & ~~\(7.69\,\mathrm{mPa\,s}\)  \\
       Bulk viscosity   & $\mu_{b}$ & ~\(1.00\,\mathrm{mPa\,s}\)\\
       Focal pressure magnitude   & $p_{\text{foc}}$ & ~\(1\,\mathrm{MPa}\)\\
       Peak velocity magnitude   & $U_0$ & the maximum axial velocity\\
       Flow Reynolds number & $Re_{\lambda}$  &  ${ \rho_m v_r \lambda}/{\mu_s}$\\
      Particle Reynolds number & $Re_s$ & ${\rho_m v_r d}/{\mu_s}$ \\
      \hline 
          \hline 
  \end{tabular}
  \label{tab:Physical parameters}
  \end{center}
\end{table}

\section{Results and discussion}\label{sec:results and discussion}

\subsection{Flow characteristics}\label{sec:Flow characteristics}

\subsubsection{Streaming flow structure}\label{sec:Flow structure}

\par
\begin{figure} 
\centering 
\includegraphics[width=8.6cm]{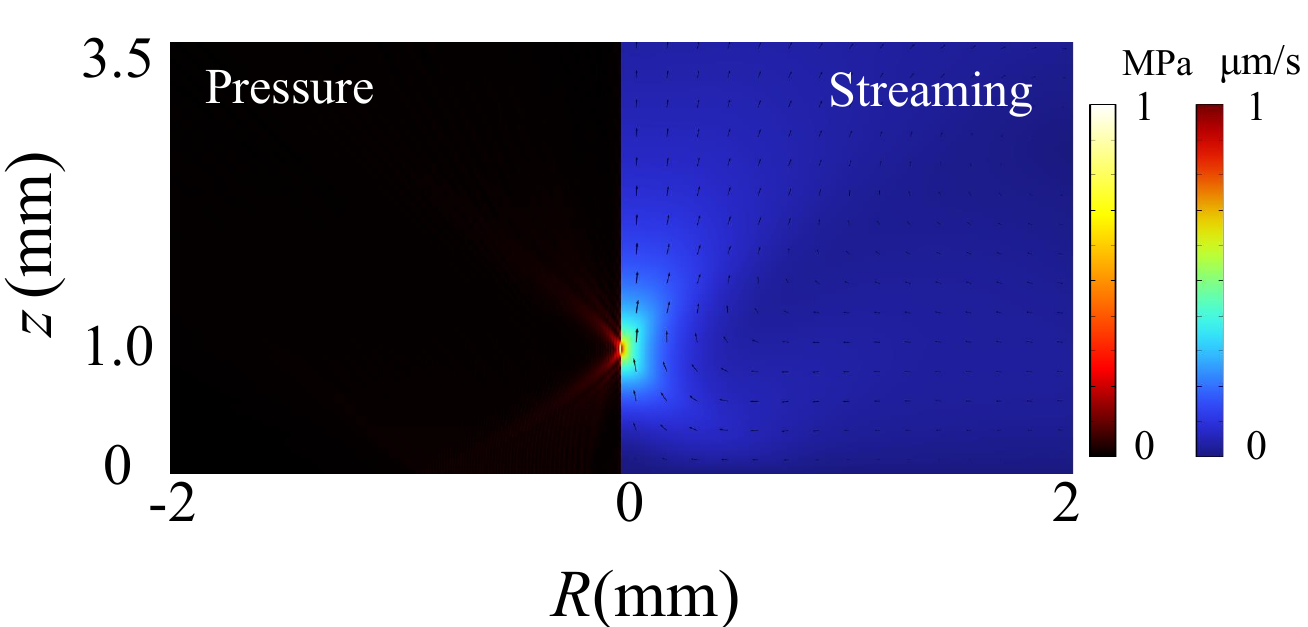}
\caption{
Finite element simulation of the acoustic pressure field (left) and the resulting acoustic streaming flow (right). 
Black arrows indicate the streaming velocity vectors. 
The transducer operates at the frequency of $f = 40$ MHz with a focal length of $h = 1$ mm, and the aperture radius of $R_A$=1.72~mm, which corresponds to $N = 26$ turns. 
The generated focal pressure amplitude is $p_{\mathrm{foc}} = 1$ MPa.  The maximum flow velocity is approximately 1 mm/s.
Both the acoustic and streaming fields exhibit axisymmetric distributions, consistent with the geometry of the system.
}
\label{Fig3: N=26 h=1mm}
\end{figure}

Owing to the axial symmetry of the acoustic system, the numerically simulated acoustic streaming field exhibits a correspondingly symmetric structure, as illustrated in~Fig.\ref{Fig3: N=26 h=1mm}. This figure highlights the intrinsic coupling between the spatial distributions of the acoustic pressure field and the induced streaming flow, both of which display strong axisymmetric characteristics. The flow field is visualized via velocity vectors.
As the focal pressure increases, the peak streaming velocity also rises, prompting a transition from creeping (viscosity-dominated) to laminar flow (inertia-influenced). This transition is characterised by the flow Reynolds number $Re_{\lambda}$, defined with respect to the acoustic wavelength $\lambda$ as the characteristic length \citep[]{friend2011microscale,dubrovski2023theory}. The definition of the flow Reynolds number $Re_{\lambda}$ is given in Table \ref{tab:Physical parameters}. By setting $Re_{\lambda} = 1$, one can derive the threshold velocity $v_{\mathrm{cr}}$ separating the two regimes:
\begin{equation}
    v_{\mathrm{cr}} = \frac{\mu_s}{\rho_m c_m} f = 0.15\, \mathrm{m/s},
\end{equation}
where $\mu_s$, $\rho_m$, and $c_m$ denote the dynamic viscosity, fluid density, and sound speed in the medium, respectively.
\par
\begin{figure}
\centering
\includegraphics[width=7.6cm]{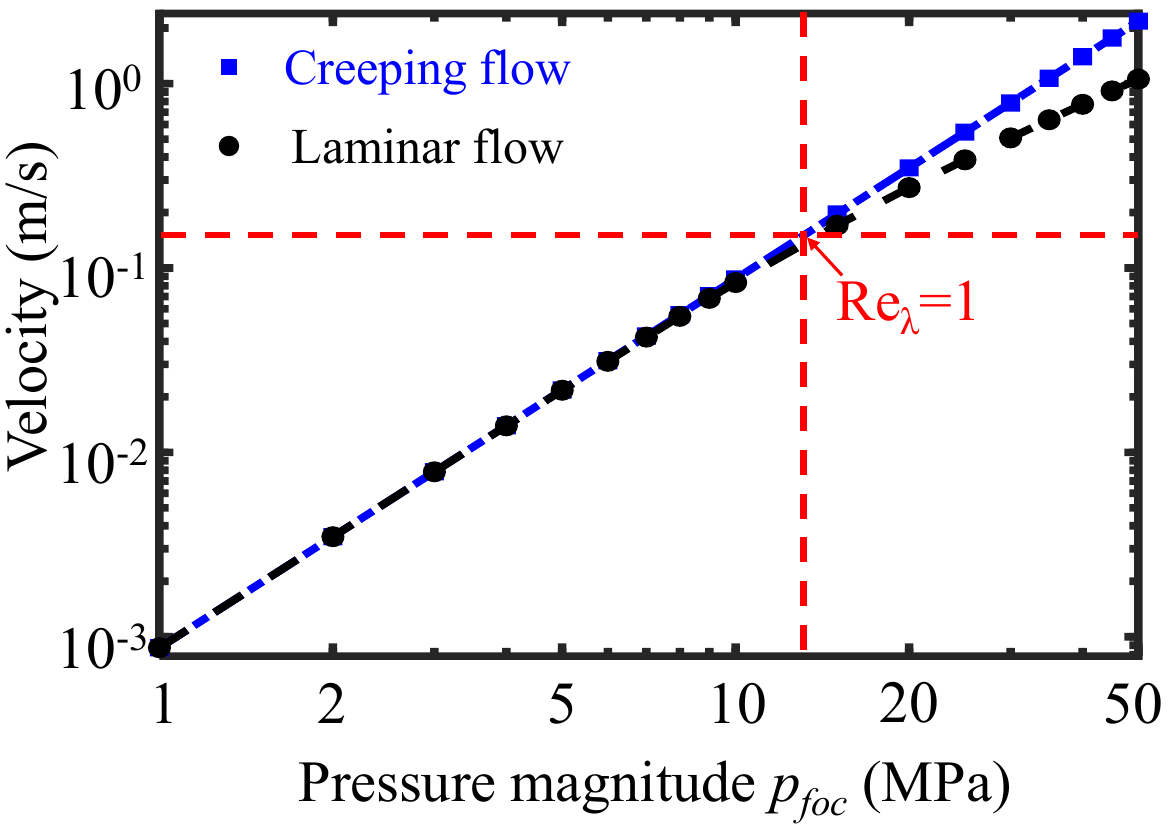}
\caption{
Log–log plot of streaming velocity versus focal acoustic pressure, comparing creeping- and laminar-flow regimes. The red dashed line marks the critical threshold $Re_{\lambda} = 1$ (critical velocity \( v_{\mathrm{cr}} = 0.15~\mathrm{m/s} \)
) and the corresponding focal pressure. The pressure magnitude is evaluated at the focal point. The calculations were conducted at an operating frequency of 40 MHz, with a focal length of $h=1~\text{mm}$ and the transducer aperture radius of $R_A$=1.72~mm (corresponding to $N$=26 turns).
}
\label{Fig: velocity versus p}
\end{figure}

Fig.\ref{Fig: velocity versus p} shows the simulated streaming velocity as a function of focal pressure amplitude, spanning both creeping and laminar flow regimes. At low pressures, where $Re_{\lambda} \ll 1$, the predicted velocities under both regimes show close agreement, indicating that viscous forces dominate and inertial contributions remain negligible. As the focal pressure increases and $Re_{\lambda}$ approaches or exceeds unity ($Re_{\lambda} \sim 1$), notable deviations arise, reflecting the growing influence of inertia on the streaming structure. Combining the critical velocity criterion with the velocity–pressure relation yields a threshold focal pressure of approximately 14~MPa, marking the onset of inertial streaming. This transition marks a fundamental change in the scaling behavior of the streaming flow. Therefore, it is essential to analyze the streaming velocity scaling laws separately within each flow regime to accurately capture the underlying dynamics—a topic explored in detail in Appendix~\ref{appA}.
\par

\subsubsection{Jetting velocity scaling law}\label{sec:scaling law}

In acoustofluidic systems, the total force acting on suspended particles results from the combined effects of acoustic radiation force and streaming-induced drag. While the radiation force can be analytically described, estimating the drag force depends on knowledge of the steady-state streaming velocity, which often requires numerical simulation due to the lack of an analytical expression. To address this, we derive a unified scaling for the jetting velocity that bridges viscous- and inertia-dominate streaming, yielding a bounded local pressure exponent $4/3<n<2$.

As discussed in Section~\ref{sec:Flow structure}, the flow field transitions from creeping to laminar as the acoustic pressure increases, corresponding to a shift in the flow Reynolds number $Re_{\lambda}$. To capture the distinct fluid dynamics across this transition, we analyse the velocity scaling behaviour within three regimes: the viscous-dominated regime ($Re_\lambda \ll 1$), the inertia-dominated regime ($Re_\lambda \gg 1$), and the transitional regime ($Re_\lambda \sim 1$). The full derivations are presented in Appendix~\ref{appA}, and here we summarize the key results.

\begin{figure}
\centering 
\includegraphics[width=8.6cm]{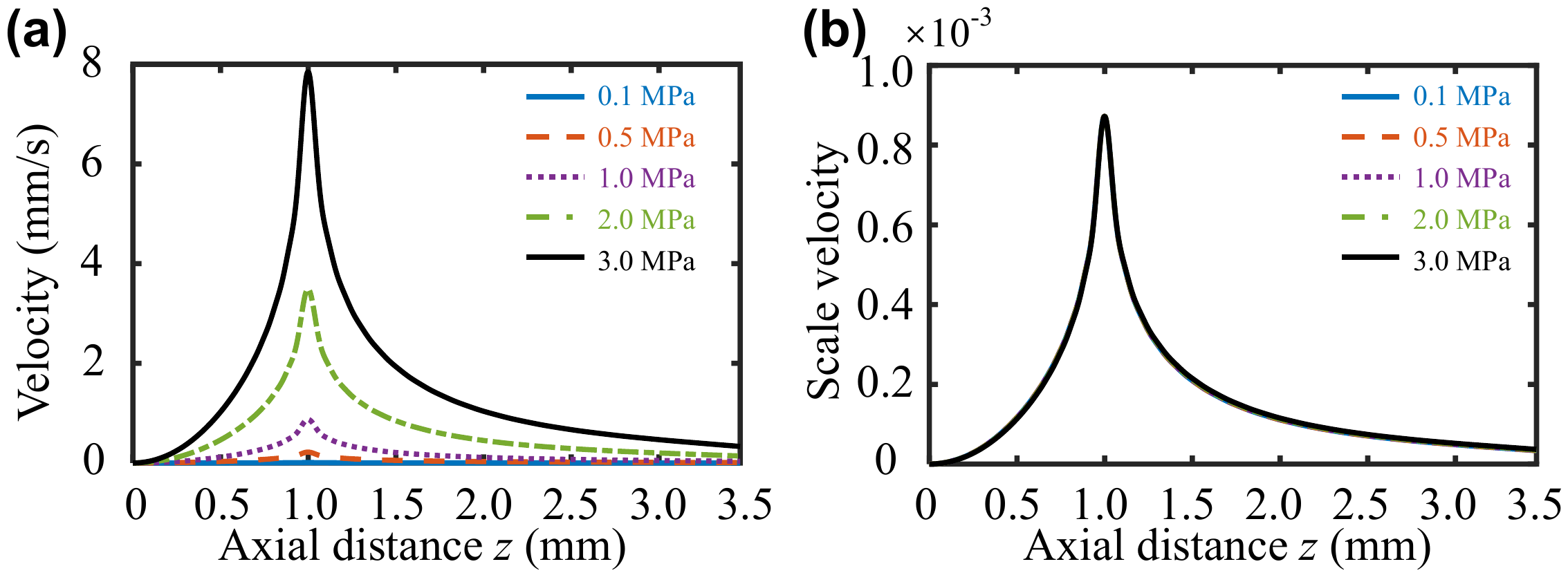}
\caption{Velocity scaling characteristics in the viscosity-dominated regime. The acoustic field operates at 40\,MHz with a focal length of 1\,mm, generated by a transducer with an aperture radius of $R_A$=1.72~mm. (a) Axial velocity profiles along the beam axis under various focal pressures. (b) Corresponding normalized velocity profiles, obtained by dividing each curve in (a) by the square of its respective focal pressure, confirming a quadratic dependence in agreement with theoretical scaling predictions.}
\label{Fig: 2 law}
\end{figure}

At low flow Reynolds number ($Re_\lambda \ll 1$), the fluid motion is governed by a balance between the acoustic body force and viscous dissipation. This yields a quadratic scaling law for the streaming velocity:
\begin{equation}
    U_0 \sim p_{\mathrm{foc}}^2.
\end{equation}
where $U_0$ denotes the peak axial streaming velocity magnitude and $p_{\rm foc}$ is the focal acoustic pressure amplitude. This scaling behaviour, initially proposed on the basis of local force-balance arguments~\citep{moudjed2014scaling}, is here derived via an energy-based global analysis. Fig.\ref{Fig: 2 law} illustrates the result: the left panel displays axial velocity profiles at various focal pressures, while the right panel demonstrates excellent collapse upon normalisation by $U_0/p_{foc}^2$, thereby confirming the quadratic pressure scaling.

As the focal pressure increases ($Re_\lambda \gg 1$), inertial effects become appreciable, particularly in the pre-focal acceleration zone. In this inertia-dominated regime, a balance emerges between acoustic forcing and convective transport, giving rise to a distinct scaling law:
\begin{equation}
    U_0 \sim p_{\mathrm{foc}}^{4/3}.
\end{equation}

\begin{figure}
\centering 
\includegraphics[width=8.6cm]{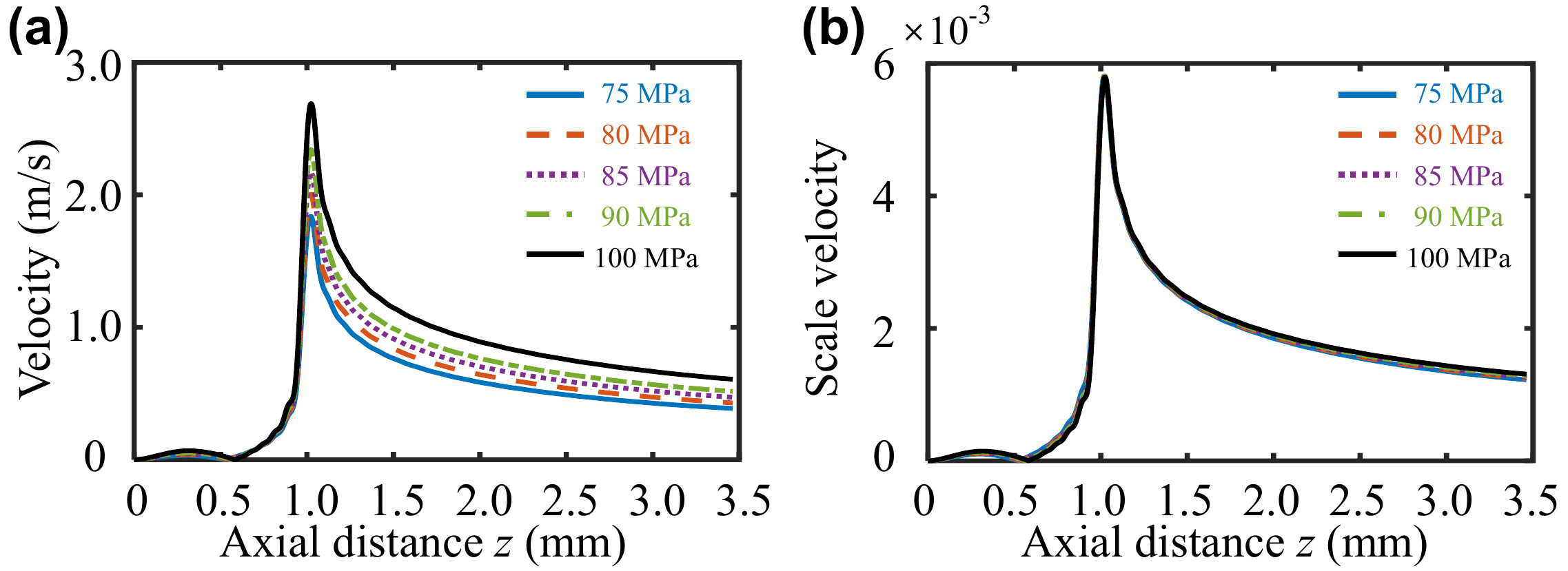}
\caption{
Velocity scaling characteristics in the inertia-dominated regime. The acoustic field operates at 40\,MHz with a focal length of 1\,mm, generated by a transducer with an aperture radius of $R_A$=1.72~mm. (a) Axial velocity profiles along the beam axis under various focal pressures. (b) Corresponding normalized velocity profiles, obtained by dividing each curve in (a) by the focal pressure raised to the power of $4/3$, confirming a sub-quadratic scaling behavior in accordance with theoretical predictions.
}
\label{Fig: 4/3 law}
\end{figure}

This 4/3 scaling has previously been reported in the literature~\citep[]{slama2019characterization}, however, we derive it here as the asymptotic limit of a unified model that smoothly bridges the viscous and inertial regimes (details see Appendix), This model not only constrains the effective pressure exponent within a bounded range, but also elucidates the influence of acoustic attenuation and beam geometry effects not captured by local scaling arguments based solely on peak velocity. As shown in~Fig.\ref{Fig: 4/3 law}, the simulation results in the high-$Re_{\lambda}$ regime closely follow the predicted power-law trend, thereby validating the theoretical scaling.

In the transitional regime ($Re_\lambda \sim 1$), viscous and inertial contributions are comparable, and the streaming velocity no longer follows a simple power law. Despite the practical significance of this transitional regime, rigorous theoretical studies on it are still notably lacking to date. 
The boundedness proof of the exponent in the transition regime is of notable significance, as it indicates that the rate of velocity change with respect to acoustic pressure gradually slows down, rather than first accelerating ($n > 2$) and then decelerating ($n \to 4/3$) . This behavior directly determines how the relative magnitude of the streaming‑induced drag force and the acoustic radiation force varies with pressure.
The lack of a rigorous analytical model hinders a unified evaluation of these two forces. To bridge this gap, we employ energy conservation and derive a cubic relation for the streaming velocity (see Appendix~\ref{appA} for details):

\begin{equation} \label{eq:U0_cardano}
U_0^3 + C_1 p_{\mathrm{foc}}^2 U_0 - D_1 p_{\mathrm{foc}}^4 = 0,
\end{equation}
which leads to a scaling exponent $n$:
\begin{equation}
n(p_{\mathrm{foc}})\;\equiv\;\frac{d\ln U_0}{d\ln p_{\mathrm{foc}}}
\;=\;2-\frac{2U_0^2}{\,3U_0^2+C_1 p_{\mathrm{foc}}^{2}\,}.
\label{eq:def-n}
\end{equation}
This result implies a transition of the exponent from $n = 2$ to $n = 4/3$ as $p_{\mathrm{foc}}$ increases. 

Since the radiation force scales as $F_{\mathrm{rad}} \propto p_{\mathrm{foc}}^2$ and the drag force scales as $F_{\mathrm{drag}} \propto p_{\mathrm{foc}}^n$ with $n < 2$ in the transition and inertial regimes, increasing acoustic pressure appears to lead to a regime where radiation force dominates. This conclusion is predicated on evaluating particle resistance using Stokes drag. However, for moderate particle Reynolds numbers (\(Re_s\sim 1\)), finite-\(Re_s\) drag correlations are no longer linear in the relative velocity, accordingly, the influence of acoustic pressure should be assessed via the trap ratio \(\Gamma = F_{\mathrm{rad}}/F_{\mathrm{drag}}\).

\subsection{The influence of transducer parameters}\label{sec:transducer}

In our previous work, we proposed a strategy for selective single-cell trapping using focused ultrasound, achieved by reversing the acoustic contrast factor through the use of biocompatible media~\citep[]{li2025reversing}. However, subsequent studies have shown that in high-frequency focused fields, the drag force arising from acoustic streaming may exceed the acoustic radiation force, thereby undermining trapping stability. This observation motivates a systematic parametric investigation of transducer design, aimed at providing qualitative guidance for performance optimisation.

\begin{figure} 
\centering 
\includegraphics[width=9cm]{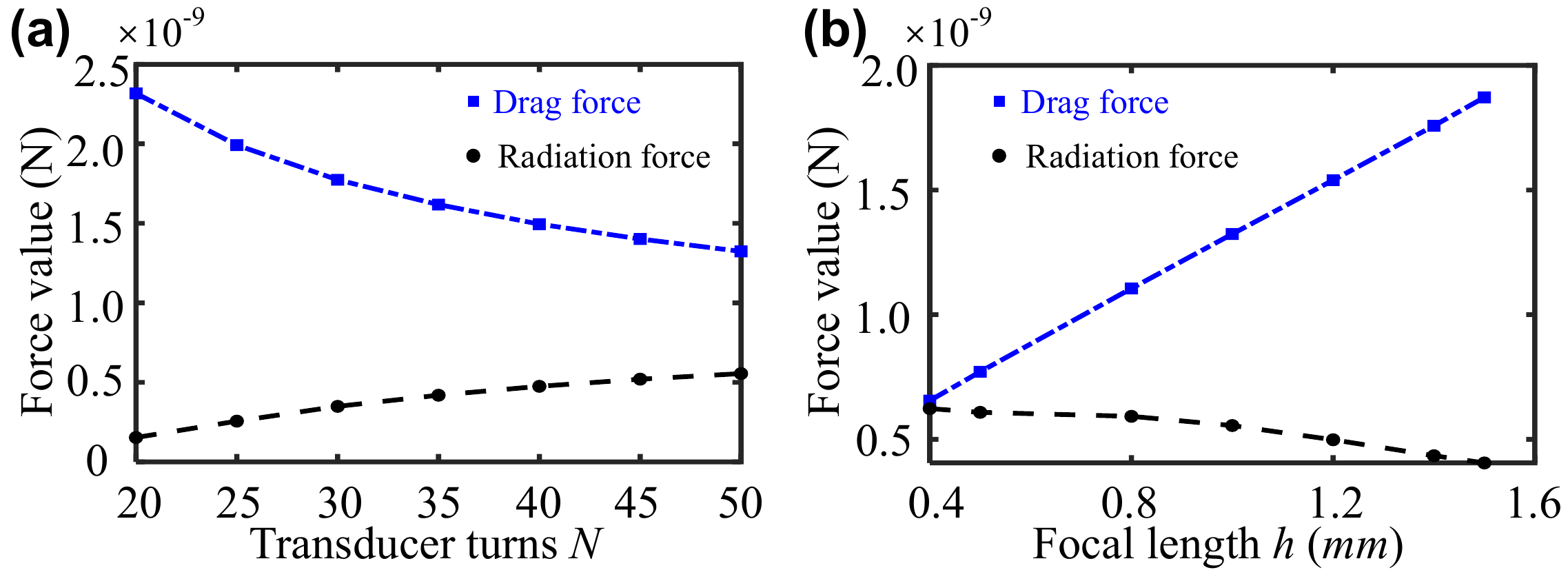} 
\caption{Effect of transducer parameters on acoustic radiation force and drag force.(a) Influence of transducer turns $N$ (corresponding to transducer aperture radius): For a fixed focal length of $h$=1 mm and a focal plane pressure of 1 MPa, the results indicate that as the number of turns ($N$) increases, the acoustic radiation force increases, while the drag force decreases.(b) Influence of transducer focal length. With the number of turns fixed at $N=50$ and the focal plane pressure maintained at 1 MPa, the results indicate that with increasing in focal length leads to a reduction in acoustic radiation force but an enhancement in drag force. It should be noted that the cell size $R$ used for the calculation corresponds to the size at which the minimum axial radiation force occurs under the given transducer parameters.}
\label{Fig: transducer parameter}
\end{figure}

To this end, we examine how the number of turns \( N \) (which governs the transducer aperture $R_A$) and the focal length \( h \) influence the acoustic radiation force and streaming-induced drag force. Since $N$ takes natural-number values, it is adopted as the primary independent variable in the parametric study. As shown in~Fig.\ref{Fig: transducer parameter}a, increasing $N$ from 20 to 50 leads to an expansion of the aperture radius from 1.46~mm to 2.71~mm.
The corresponding simulation results are summarized in~Fig.\ref{Fig: transducer parameter}. As indicated by the trend lines, increasing the number of turns (i.e., enlarging the aperture) enhances the acoustic radiation force while reducing the streaming-induced drag. Conversely, increasing the focal length weakens the radiation force and amplifies the drag. For consistency, the particle size in all simulations is selected to correspond to the minimum axial radiation force under each transducer configuration, further computational details can be found in reference \citep[]{li2025reversing}.

It is important to emphasize that all simulations were conducted under the constraint of a fixed focal sound pressure amplitude $p_{\rm foc}$=1~MPa, thereby isolating the effects of geometric parameters from variations in acoustic intensity. From a theoretical standpoint, the observed trends in streaming velocity are qualitatively consistent with the velocity scaling laws discussed in Appendix~\ref{appA}, which predict how viscous flow depends on transducer geometry.

\subsection{The influence of acoustic pressure}\label{sec:pressure}
\begin{figure}
\centering
\includegraphics[width=8.6cm]{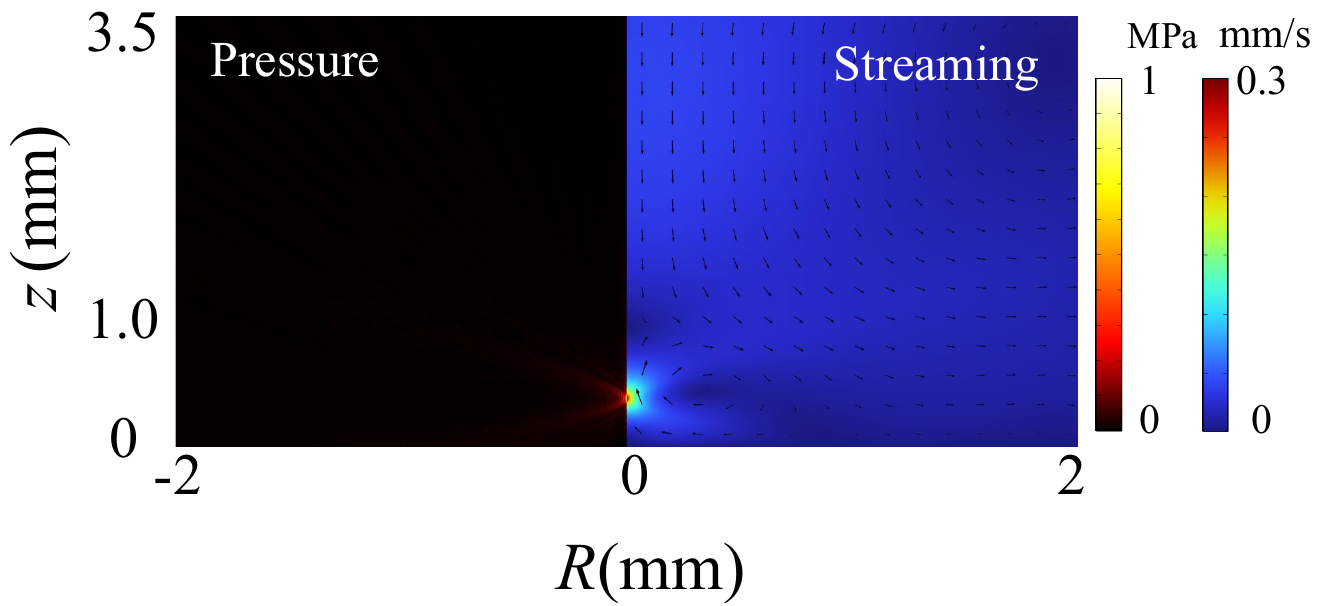}
\caption{
Finite element simulation of the acoustic pressure field and the induced streaming flow for a large-aperture transducer configuration. 
The setup is identical to that in~Fig.\ref{Fig3: N=26 h=1mm}, except with a shorter focal length $h = 0.4$ mm and a larger number of turns $N = 50$ turns (corresponding aperture radius $R_A$=2.26~mm). The maximum flow velocity is approximately 0.3 mm/s. Notably, the focused field configuration results in a backflow phenomenon downstream of the focal region.
}
\label{Fig: large aperture angle}
\end{figure}

\begin{figure}
\centering
\includegraphics[width=8.6cm]{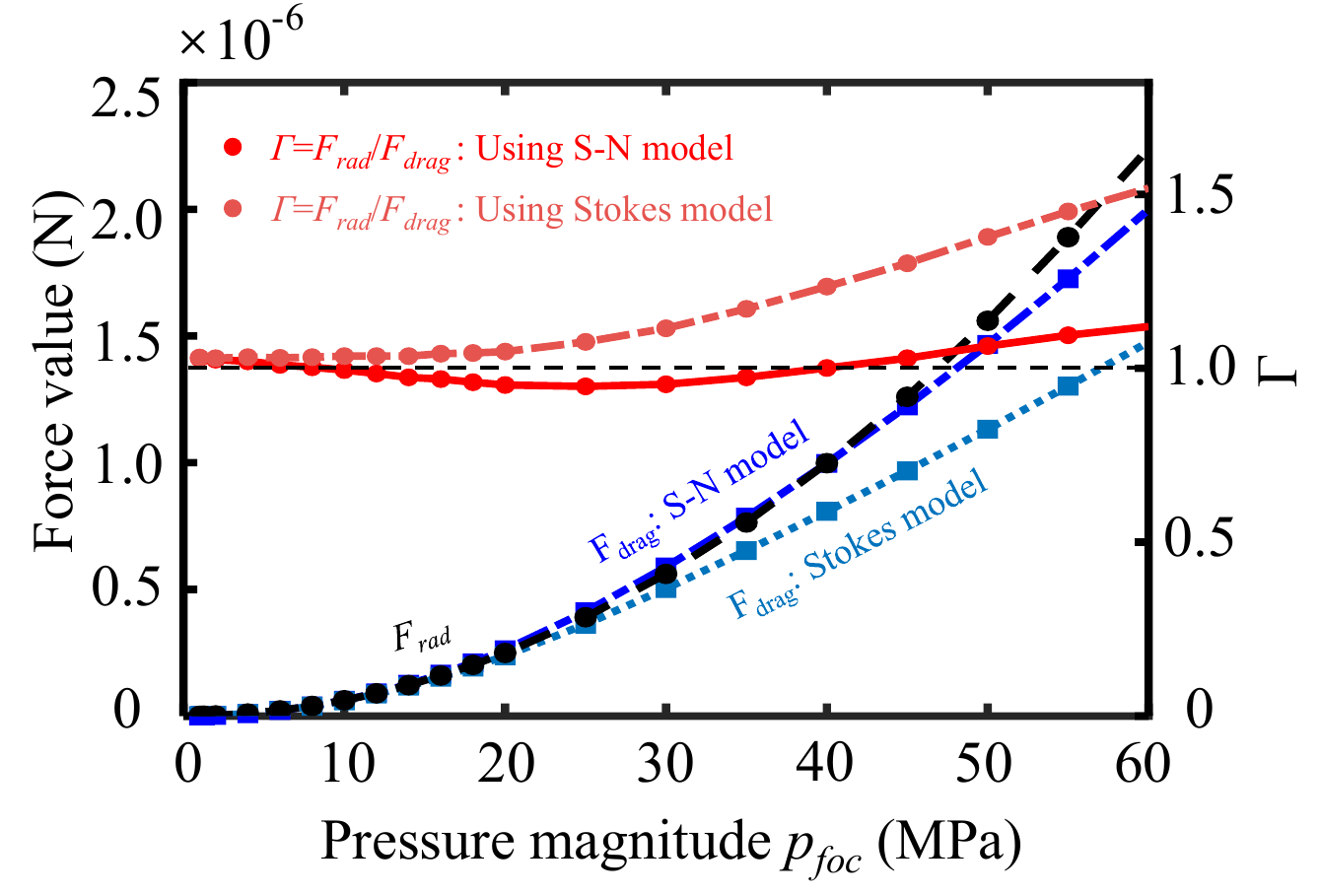}
\caption{Radiation force, drag force and the trap ratio 
$\Gamma = F_{\rm rad}/F_{\rm drag}$ 
as functions of the focal pressure magnitude $p_{\rm foc}$ for a particle of radius $R=0.38 \lambda$ suspended in a 60\%~(v/v) iodixanol solution at $40\,{\rm MHz}$. Left axis: acoustic radiation force $F_{\rm rad}$ (black, dashed), drag force predicted by the Schiller–Naumann (S–N) model $F_{\rm drag}$ (blue, dot–dashed), and classical Stokes drag (light blue, dotted). Right axis: trap ratio $\Gamma$ computed using the S–N drag model (red, solid) and the Stokes drag model (light red, solid). The Stokes-based $\Gamma$ increases monotonically with $p_{\rm foc}$ whereas the S–N-based $\Gamma$ exhibits a non-monotonic trend.
}
\label{Fig: Pressure}
\end{figure}

Based on the transducer parameter study, we select a representative configuration consisting of 50 turns (corresponding aperture radius $R_A$=2.26~mm) and a focal length of $0.4\,\text{mm}$ to examine the influence of sound pressure. Moreover, as shown in~Fig.\ref{Fig: large aperture angle}, the large aperture induces a flow reversal downstream of the acoustic focus, which reduces the peak streaming velocity and thereby mitigates the hydrodynamic drag. Both the acoustic radiation force and the streaming-induced drag force are calculated as functions of the focal pressure. As shown in ~Fig.\ref{Fig: Pressure}, varying the focal pressure amplitude modulates the relative magnitudes of the acoustic radiation force and the streaming-induced drag. Under the classical Stokes drag model (Eq.\eqref{Eq:Stokes Fdrag}), the trap ratio $\Gamma$ increases monotonically with focal pressure $p_{\rm foc}$, suggesting that stronger pressures invariably favour radiation force dominance and thus promote particle trapping. However, when the more refined S–N drag model (Eq.\eqref{Eq:Fdrag}) is employed, the trap ratio exhibits a non-monotonic dependence on pressure. This observation highlights that increasing the focal pressure does not necessarily lead to improved trapping stability.
To interpret the non-monotonic variation of the trap ratio $\Gamma \equiv F_{\rm rad}/F_{\rm drag}$ observed in~Fig.\ref{Fig: Pressure}, we invoke scaling arguments and introduce a logarithmic-slope criterion to quantify the competing trends. The radiation force maintains a quadratic dependence on focal pressure, $F_{\rm rad}\propto p_{\rm foc}^{2}$. Defining the effective pressure exponent for the drag force as $m_{\rm eff}(p_{\rm foc}) \equiv {\rm d}\ln F_{\rm drag}/{\rm d}\ln p_{\rm foc}$, it follows that
\begin{equation}
    \frac{{\rm d}\ln \Gamma}{{\rm d}\ln p_{\rm foc}} \;=\; 2 - m_{\rm eff}(p_{\rm foc}).
\end{equation}

Under the creeping-flow hypothesis with Stokes drag, increasing the pressure amplitude enhances the dominance of the radiation force, thereby improving trapping robustness~\citep[]{li2019axial}. However, our results show that this trend need not be monotonic: as acoustic pressure increases, streaming (and thus streaming-induced drag) strengthens, while finite-$Re_s$ corrections also increase drag as the particle Reynolds number $Re_s$ grows. This is particularly important for Mie particles in laminar flow within acoustic trapping. The key diagnostic is the logarithmic-slope condition $m_{\rm eff}(p^{\star})=2$. At this turning point $p^{\star}$, the drag growth exactly offsets the $p_{foc}^2$ scaling of the radiation force, causing $\Gamma$ to decrease for $m_{\rm eff}>2$ and increases for $m_{\rm eff}<2$. 
Hence the monotonicity of $\Gamma$ is governed entirely by whether $m_{\rm eff}$ is larger than, equal to, or smaller than $2$. Initially, at low pressure (viscous-dominated regime), second-order streaming scales as $U_0 \propto p_{\rm foc}^{2}$, so $m_{\rm eff}\approx 2$ and $\Gamma$ is nearly constant. As $p_{\rm foc}$ rises, finite particle Reynolds number $Re_s$ drag corrections can drive $m_{\rm eff}>2$, causing $\Gamma$ to decrease. Eventually, when the flow enters an inertia-modulated regime ($U_0 \propto p_{\rm foc}^{\,n}$,with $n<2$), $m_{\rm eff}$ falls below and crosses $2$, and $\Gamma$ turns upward. The turning point is set by $m_{\rm eff}(p^{\star})=2$ and does not require reaching the high-pressure asymptote (e.g.\ $n=4/3$); under the present parameters this crossing already occurs at moderate pressures.

It should be noted that the present analysis temporarily relaxes constraints imposed by the mechanical index (MI)~\citep{sen2015mechanical_index, aium2021biological_effects}. If the influence of MI is to be considered, further optimization can be performed within the research framework and theoretical model developed in this work. The primary objective of this study is to elucidate the dominant physical mechanisms governing 3D cell trapping in high-frequency focused acoustic tweezers, and to identify general strategies for device and parameter optimization. The insights gained here may also inform the broader design of acoustic tweezers in other operating regimes.

\section{Conclusions}
\label{sec:conclusions}
We developed a unified theoretical--numerical framework for single-beam acoustic tweezers based on focused beam that combined ARF with drag force induced by the ABS, and enables quantitative comparison of radiation and drag over a broad parameter range. The peak streaming flow velocity obeys a bounded pressure scaling with local pressure exponent \(4/3<n<2\), approaching \(n=2\) in the viscous limit and \(n=4/3\) when inertial transport dominates, a cubic balance links these limits and provides a practical estimate across the crossover. This resolves prior ambiguity in how streaming responds to pressure in practical devices. Crucially, we find that the trap ratio \(\Gamma=F_{\rm rad}/F_{\rm drag}\) does not increase monotonically with pressure: a slope criterion $m_{\rm eff}=2$ governs the monotonicity of $\Gamma$, particularly for Mie-size particles at moderate Reynolds number. This explains the emergence of performance-optimal pressure windows. Moreover, geometric parameters such as aperture and focal length act as effective design levers that reshape the streaming jet—larger apertures induce downstream recirculation that weakens axial drag and favors trapping. The framework is broadly applicable to homogeneous Newtonian fluids at low acoustic Mach number, and may be extended to include thermal and compositional effects relevant to biological suspensions and inhomogeneous media. In particular, thermoviscous corrections to the radiation force term, including near particle boundary layer dissipation and related absorbed power effects, are not resolved explicitly in the present work and remain an important direction for future study.

\begin{acknowledgments}
 Z. Gong thanks for the support from the National Natural Science Foundation of China (24Z990200542 and No. 12504522), the XIAOMI Foundation, and the Shanghai Jiao Tong University [2030 Initiative, AI for Engineering Initiative, and the startup funding (WH220401017, WH22040121)].
\end{acknowledgments}

\section*{\label{sec 5}AUTHOR DECLARATIONS}
\textbf{Conflict of Interest} \\
The authors have no conflicts to disclose.

\section*{\label{sec 6}DATA AVAILABILITY}
The data that support the findings of this study are available from the corresponding author upon reasonable request.
\appendix

\section{Scaling Law Analysis}\label{appA}
The derivation of the scaling laws begins with the kinetic energy equation. By analysing the balance among the dominant terms, the corresponding regime-specific scalings are obtained:
\begin{equation}
\begin{aligned}
&\iiint_V \rho\!\left(
    \frac{\partial e_c}{\partial t}
    + \boldsymbol{v}\cdot\nabla e_c
\right) dV
= -2\mu_s \iiint_V \boldsymbol{D}:\boldsymbol{D}\, dV
\\[3pt]
&\phantom{=}+ \iint_{\partial V} \boldsymbol{\sigma}\cdot\boldsymbol v\cdot \boldsymbol n\, dS
  + \iiint_V \boldsymbol{f}_s\cdot\boldsymbol v\, dV .
\end{aligned}
\end{equation}

where $e_c=\tfrac12\,\boldsymbol{v}\!\cdot\!\boldsymbol{v}$ is the kinetic energy density, and ${{\boldsymbol{\sigma}}} = -p \boldsymbol{I} + 2\mu_s \boldsymbol{D}$ is the stress tensor in the incompressible fluid~\citep[]{daru2024high}.
\subsection{Scaling law in the viscosity–dominated regime}
\label{sec:app:viscous-scaling}

At sufficiently low focal pressures ($Re_{\lambda}\ll 1$), the streaming is governed by a balance between the acoustic power input and viscous dissipation. The corresponding energy budget over the control volume is expressed as
\begin{equation}
\iiint_V \boldsymbol{f}_s\!\cdot\!\boldsymbol{v}\,dV
\;\sim\;
\iiint_V 2\mu_s\,\boldsymbol{D}:\boldsymbol{D}\,dV,
\label{A1}
\end{equation}
where $\boldsymbol{D}$ is the rate of strain tensor and $\boldsymbol{f}_s$ denotes the acoustic body force. Estimating the left–hand side by the acoustic power over a cross–section of area
$\pi R_{\rm beam}^2$ and the right–hand side by a dissipation rate based on a shear
scale $R_{\rm jet}$ in a control volume of length $L$ gives
\begin{equation}
\begin{aligned}
U_0 \,\pi R_{\rm beam}^2 \int_0^L f_s(z)\,dz
&\;\sim\;
2\mu_s\,\Big(\frac{U_0}{R_{\rm jet}}\Big)^2 \,\pi R_{\rm jet}^2 L
\\
&\;\sim\;
2\mu_s\,U_0^2\,\pi L .
\end{aligned}
\label{Eq:A2}
\end{equation}

where the $R_{beam}$ and $R_{jet}$ represent the radius of the acoustic beam and the resulting streaming jet, respectively.

For convenience we introduce the focal pressure magnitude envelope:
\begin{equation}
p(z)=p_{\rm foc}\,G(z),\qquad G(h)=1,
\label{A4}
\end{equation}
Here $G(z)$ denotes a dimensionless axial envelope with maximum value $G_{\max}=1$.
Within the framework of linear acoustics, the normalised pressure profiles corresponding to different driving amplitudes collapse onto a single universal curve. This collapse indicates that, under the assumption of linear propagation, the use of a dimensionless acoustic envelope function
$G(z)$ provides a consistent and scalable description of the axial pressure field, independent of the absolute pressure amplitude.

The streaming body force is expressed in the standard Eckart form:
\begin{equation}
f_s(z)\;\sim\;\alpha\,\frac{I(z)}{c_m}
\;\sim\;\alpha\,\frac{I_{\rm foc}}{c_m}\,G^2(z)\,e^{-2z/L_a},
\label{Eq:body force}
\end{equation}
where $L_a=\alpha^{-1}$ is the attenuation length and $\alpha$ is the attenuation coefficient.

Substituting Eq.\eqref{Eq:body force} into the balance Eq.\eqref{Eq:A2}, together with the cross-sectional and volumetric estimates therein, leads to the scaling relation for the streaming velocity in the viscosity-dominated regime.
\begin{equation}
U_0
\;\sim\;
\frac{\alpha\,I_{\rm foc}\, R_{\rm beam}^2}{2\mu_s\, L\,c_m}
\int_0^{L} G^2(z)\,e^{-2z/L_a}\,dz
\;\sim\; p_{\rm foc}^{\,2}.
\label{A5}
\end{equation}

Using the diffraction estimate $R_{\rm beam}=h\lambda/R_A$, we obtain the following refined expression.
\begin{equation}
U_0
\;\sim\;
\frac{\alpha\,I_{\rm foc}\,h^{2}\lambda^{2}}{2\mu_s L c_m R_A^{2}}
\int_0^{L} G^2(z)\,e^{-2z/L_a}\,dz
\;\sim\; p_{\rm foc}^{\,2}.
\label{A6}
\end{equation}

Before using Eq.\eqref{A5}–Eq.\eqref{A6}, we first verify that the right-hand sides possess the correct dimensions of velocity and that the focal pressure magnitude $p_{\rm foc}$ is the sole term carrying pressure dependence. In SI units, this yields:
\begin{equation}
\begin{aligned}
[I_{\rm foc}] &= \mathrm{kg\,s^{-3}}, &
[\alpha]      &= \mathrm{m^{-1}},      &
[R_{\rm beam}]&= \mathrm{m},\\
[\mu_s]       &= \mathrm{kg\,m^{-1}\,s^{-1}}, &
[L]           &= \mathrm{m},            &
[c_m]         &= \mathrm{m\,s^{-1}}.
\end{aligned}
\end{equation}
and \(G\) is dimensionless, so the integral
\(
\big[\!\int_0^L G^2 e^{-2z/L_a}dz\big]={\rm m}.
\)
Accordingly, the right-hand side of Eq.\eqref{A5} retains the units of velocity:
\begin{equation}
\begin{aligned}
\frac{\alpha\,I_{\rm foc}\,R_{\rm beam}^2\ \int_0^L \!\cdots dz}
     {2\,\mu_s\,L\,c_m}
&=
\frac{ (\mathrm{m^{-1}})(\mathrm{kg\,s^{-3}})(\mathrm{m^2})(\mathrm{m}) }
     { (\mathrm{kg\,m^{-1}\,s^{-1}})(\mathrm{m})(\mathrm{m\,s^{-1}}) }
\\[6pt]
&=
\frac{ \mathrm{kg\,m^{2}\,s^{-3}} }
     { \mathrm{kg\,m\,s^{-2}} }
= \mathrm{m\,s^{-1}} .
\end{aligned}
\end{equation}

In Eq.\eqref{A6}, we adopt the diffraction-based estimate
\(R_{\rm beam}=h\,\lambda/R_A\) where both the focal length $h$ and the aperture radius $R_A$ have dimensions of length $\rm m$. The resulting substitution,
\[
\left[\frac{h^2\lambda^2}{R_A^2}\right]
=\frac{({\rm m^2})({\rm m^2})}{({\rm m^2})}
={\rm m^2},
\]
confirms that replacing $R_{\rm beam}^2$ with $h^2\lambda^2/R_A^2$ preserves the dimensional consistency of Eq.\eqref{A6}, whose right-hand side retains the expected units of velocity (\({\rm m\,s^{-1}}\)).  

To isolate the dependence on pressure, we substitute
\(I_{\rm foc}=p_{\rm foc}^2/(2\rho_m c_m)\) where the focal pressure has units
\([p_{\rm foc}]={\rm kg\,m^{-1}\,s^{-2}}\). All other parameters—namely, the attenuation coefficient~$\alpha$, viscosity~$\mu_s$, density~$\rho_m$, sound speed~$c_m$, focal length~$h$, wavelength~$\lambda$, aperture radius ~$R_A$, control volume length~$L$, and the integral~$\int_0^L\!\cdots dz$-are considered fixed geometric or material constants for a given device and medium. Consequently, the streaming velocity scales as
\[
U_0\ \propto\ I_{\rm foc}\ \propto\ p_{\rm foc}^{\,2},
\]
and the low-pressure exponent is \(n=2\).

\subsection{Scaling law in the inertia–dominated regime}
\label{sec:app:inertia-scaling}

At sufficiently high focal pressures ($Re_{\lambda} \gg 1$), inertial transport balances the acoustic power input.
The corresponding energy transfer within the control volume is given by:
\begin{subequations}\label{B2}
\begin{align}
\rho_m U_0^2 &\sim
\frac{\alpha R_{\rm beam}^2}{c_m w_{\rm jet}^2}
\int_0^L I(z)\,e^{-2z/L_a}\,dz, \\
I(z) &= I_{\rm foc}\,G^2(z)\,e^{-2z/L_a}.
\end{align}
\end{subequations}

At the axial location where the streaming velocity attains its maximum, the axial gradient vanishes, $\partial u/\partial z=0$, and the viscous term is balanced by the acoustic body force,
\begin{equation}
\frac{2\alpha\,I_{\rm foc}}{c_m} \;\sim\; \mu_s\,\frac{U_0}{R_{\rm jet}^2}
\quad\Rightarrow\quad
R_{\rm jet}^2 \;\sim\; \frac{\mu_s\,c_m}{2\alpha\,I_{\rm foc}}\,U_0.
\label{B3}
\end{equation}
Eliminating $R_{\rm jet}$ between Eq.\eqref{B2}–Eq.\eqref{B3} yields a cubic relation for the peak streaming velocity $U_0$:
\begin{equation}
\rho_m\,U_0^3 \;\sim\; \frac{2\alpha^2 R_{\rm beam}^2}{\mu_s\,c_m^2}\,I_{\rm foc}^2\!
\int_0^L G^2(z)\,e^{-2z/L_a}\,dz.
\label{B4}
\end{equation}
Using $I_{\rm foc}=p_{\rm foc}^2/(2\rho_m c_m)$ and $R_{\rm beam}=h\,\lambda/R_A$ , then yields the scaling law for the inertia-dominated regime
\begin{equation}
U_0 \;\sim\;
\left[
\frac{\alpha^2 h^2\lambda^2}{2\,\rho_m^{3}\mu_s\,c_m^{4}R_A^{2}}
\int_0^L G^2(z)\,e^{-2z/L_a}\,dz
\right]^{\!1/3} p_{\rm foc}^{\,4/3}.
\label{B5}
\end{equation}

In SI units, $[\alpha]={\rm m^{-1}}$, $[\lambda]={\rm m}$, $[\rho_m]={\rm kg\,m^{-3}}$,
$[\mu_s]={\rm kg\,m^{-1}\,s^{-1}}$, $[c_m]={\rm m\,s^{-1}}$, and
$\big[\!\int_0^L\!G^2 e^{-2z/L_a}dz\big]={\rm m}$. The bracket in Eq.\eqref{B5} thus has units
\begin{multline}
\frac{[\alpha]^2[h]^2[\lambda]^2\,[\!\int dz]}
     {[\rho_m]^3[\mu_s][c_m]^4[R_A]^2}
\\
\shoveleft{%
= \frac{\mathrm{m^{-2}}\cdot \mathrm{m^2}\cdot \mathrm{m^2}\cdot \mathrm{m}}
        {(\mathrm{kg\,m^{-3}})^3\cdot \mathrm{kg\,m^{-1}\,s^{-1}}
         \cdot (\mathrm{m\,s^{-1}})^4\cdot \mathrm{m^2}}
  = \frac{\mathrm{m^7\,s^5}}{\mathrm{kg^4}}%
}
\end{multline}

Taking the one-third power of this expression yields dimensions of ${\rm m^{7/3}s^{5/3}kg^{-4/3}}$, and multiplying by
$[p_{\rm foc}]^{4/3}={\rm (kg\,m^{-1}\,s^{-2})^{4/3}}={\rm kg^{4/3}m^{-4/3}s^{-8/3}}$
then gives ${\rm m\,s^{-1}}$, which confirms that the right-hand side of Eq.\eqref{B5} has the correct physical dimension of a velocity. Therefore, the inertia-dominated scaling $U_0\propto p_{\rm foc}^{4/3}$ is dimensionally consistent.

\subsection{Scaling law in the transitional regime}
\label{sec:app:transitional-scaling}

In the transitional regime ($Re_{\lambda} \sim 1$), acoustic body force, inertial transport, and viscous dissipation contribute at comparable orders. The corresponding control-volume balance for the steady streaming field is therefore given by:
\begin{equation}
\begin{aligned}
    &\iiint_V \rho_m\,\boldsymbol{v}\cdot\nabla e_c \, dV 
    \sim
    \iiint_V -2\mu_s\,\boldsymbol{D}:\boldsymbol{D}\, dV \\
    &\quad+
    \iiint_V \boldsymbol{f}_s\cdot\boldsymbol{v}\, dV,
\end{aligned}
\label{eq:app:energy-balance}
\end{equation}
where $e_c=\tfrac12\,\boldsymbol{v}\!\cdot\!\boldsymbol{v}$ is the kinetic–energy density, $\boldsymbol{D}$ is the rate of strain tensor, and $\boldsymbol{f}_s$ the acoustic body force. By introducing a representative axial envelope \(G(z)\), a beam radius \(R_{\text{beam}}\), and a control–volume length \(L\) along the propagation direction, (\ref{eq:app:energy-balance}) reduces to a cubic balance for the characteristic (peak) streaming speed \(U_0\):
\begin{equation}
U_0^3 \;+\; 
\underbrace{\frac{4L\,\alpha\, I_{\text{foc}}}{\rho_m c_m}}_{C}\,U_0
\;-\;
\underbrace{\frac{2 \alpha^2 I_{\text{foc}}^2 R_{\text{beam}}^2}{\rho_m \mu_s\, c_m^2}
\int_{0}^{L} G^2(z)\, e^{-2z/L_a}\, dz}_{D}
= 0.
\label{eq:app:cubic-intensity}
\end{equation}
Here, \(I_{\text{foc}}\) denotes the focal intensity, and \(L_a\) is the attenuation length. The term \(C\,U_0\) represents the viscous contribution, while \(D\) measures the acoustic driving that feeds the inertial transport.

Let $[U_0]={\rm m\,s^{-1}}$, so $[U_0^3]={\rm m^3\,s^{-3}}$. The envelope $G$ is dimensionless, hence
$\big[\!\int_0^L G^2 e^{-2z/L_a}\,dz\big]={\rm m}$. Using
$[I_{\rm foc}]={\rm kg\,s^{-3}}$, $[\alpha]={\rm m^{-1}}$,
$[\rho_m]={\rm kg\,m^{-3}}$, $[\mu_s]={\rm kg\,m^{-1}\,s^{-1}}$, $[c_m]={\rm m\,s^{-1}}$,
$[R_{\rm beam}]={\rm m}$, we obtain
\[
[C]=\frac{[L][\alpha][I_{\rm foc}]}{[\rho_m][c_m]}
=\frac{({\rm m})({\rm m^{-1}})({\rm kg\,s^{-3}})}{({\rm kg\,m^{-3}})({\rm m\,s^{-1}})}
={\rm m^2\,s^{-2}}=[U_0]^2,
\]
thus $[C\,U_0]={\rm m^3\,s^{-3}}$.
For the driving term,
\[
\begin{aligned}
[D]
&=\frac{[\alpha]^2[I_{\rm foc}]^2[R_{\rm beam}]^2}{[\rho_m][\mu_s][c_m]^2}
\Big[\!\int_0^L\!\cdots dz\Big] \\[4pt]
&=\frac{({\rm m^{-2}})\,({\rm kg^2\,s^{-6}})\,({\rm m^2})}
{({\rm kg\,m^{-3}})\,({\rm kg\,m^{-1}\,s^{-1}})\,({\rm m^2\,s^{-2}})}
\,({\rm m})
={\rm m^3\,s^{-3}} .
\end{aligned}
\]

Hence all three terms in Eq.\eqref{eq:app:cubic-intensity} carry units ${\rm m^3\,s^{-3}}$:
$[U_0^3]=[C\,U_0]=[D]$, confirming the dimensional correctness of the cubic balance.

Using the standard relation \(I_{\text{foc}}=p_{\text{foc}}^2/(2\rho_0 c_0)\), the cubic relation may be re-expressed directly in terms of the focal pressure as
\begin{equation}
U_0^3 \;+\; C_1\,p_{\mathrm{foc}}^2\,U_0 \;-\; D_1\,p_{\mathrm{foc}}^4 \;= 0,
\qquad
\label{eq:app:cubic-pressure}
\end{equation}
with $C_1={2L\,\alpha}/{\rho_m^2 c_m^2}$, $D_1={\alpha^2 R_{\text{beam}}^2}/({2\rho_m^{3}\mu_s\, c_m^{4}}\,
\int_{0}^{L} G^2(z)\,e^{-2z/L_a}\,dz)$ for a fixed device and medium.

Let $F(U,p)\equiv U^3+C_1p^2U-D_1p^4$ denote the left–hand side of Eq.\eqref{eq:app:cubic-pressure}.
This expression may be cast into the standard cubic form $x^3+rx+q=0$ with
\[
r=C_1 p_{\mathrm{foc}}^{2}>0,\qquad q=-D_1 p_{\mathrm{foc}}^{4}<0,
\]
The discriminant
\[
\Delta_{\mathrm{cub}}=-4r^{3}-27q^{2}
=-\,4C_1^{3}p_{\mathrm{foc}}^{6}-27D_1^{2}p_{\mathrm{foc}}^{8}\;<\;0
\]
indicates that Eq.\eqref{eq:app:cubic-pressure} admits a single real root, denoted $U_0(p_{\mathrm{foc}})$. To show that this root is strictly positive, we argue by contradiction. if $U_0\le 0$, then since
$\partial_U F(U,p)=3U^2+C_1p^2>0$ for all $U$ (strictly increasing in $U$),
\[
0=F(U_0,p_{\mathrm{foc}})\;\le\;F(0,p_{\mathrm{foc}})=-D_1p_{\mathrm{foc}}^4<0,
\]
which is a contradiction. Hence $U_0(p_{\mathrm{foc}})>0$. Furthermore, since
$\partial_UF(U_0(p_{\mathrm{foc}}),p_{\mathrm{foc}})=3U_0^2+C_1p_{\mathrm{foc}}^2\neq 0$, so the root
is nondegenerate. By the implicit function theorem, $U_0$ depends smoothly on $p_{\mathrm{foc}}$ for
$p_{\mathrm{foc}}>0$. This smoothness ensures that the local pressure exponent is well defined and justifies implicit differentiation of Eq.\eqref{eq:app:cubic-pressure} in what follows.
With $U_0$ smooth in $p_{\mathrm{foc}}$, define the local scaling exponent
\begin{equation}
n(p_{\mathrm{foc}})\;\equiv\;\frac{d\ln U_0}{d\ln p_{\mathrm{foc}}}
\;=\;\frac{p_{\mathrm{foc}}}{U_0}\,\frac{dU_0}{dp_{\mathrm{foc}}}.
\label{eq:app:def-n}
\end{equation}
Implicit differentiation of Eq.\eqref{eq:app:cubic-pressure}, followed by elimination of the term $D_1p_{\mathrm{foc}}^4$
using Eq.\eqref{eq:app:cubic-pressure} itself, yields the following expression for the local pressure exponent:
\begin{equation}
n(p_{\mathrm{foc}})\;=\;\frac{2\bigl(2U_0^2+C_1 p_{\mathrm{foc}}^{2}\bigr)}{3U_0^2+C_1 p_{\mathrm{foc}}^{2}}
\;=\;2-\frac{2U_0^2}{\,3U_0^2+C_1 p_{\mathrm{foc}}^{2}\,}.
\label{eq:app:n-closed}
\end{equation}
Two immediate consequences of Eq.\eqref{eq:app:n-closed} are
\begin{equation}
\begin{aligned}
n-2 &= -\frac{2U_0^2}{3U_0^2 + C_1 p_{\mathrm{foc}}^{2}} < 0 ,\\[4pt]
n-\frac{4}{3}
&= \frac{2 C_1 p_{\mathrm{foc}}^{2}}
{3\,(3U_0^2 + C_1 p_{\mathrm{foc}}^{2})} > 0 .
\end{aligned}
\label{eq:app:n-bounds}
\end{equation}

and hence, for all $p_{\mathrm{foc}}>0$,
\begin{equation}
 \frac{4}{3}\;<\;n(p_{\mathrm{foc}})\;<\;2.
\label{eq:app:n-range}
\end{equation}

Finally, balancing the terms in Eq.\eqref{eq:app:cubic-pressure} recovers the canonical limiting behaviours that define the pressure-scaling bounds and complete the causal chain from the cubic relation to the local exponent: at low focal pressure, a viscous-dominated balance yields $U_0\sim p_{\mathrm{foc}}^{2}$ so $n\to 2$, whereas at high pressure, an inertia-influenced balance yields $U_0\sim p_{\mathrm{foc}}^{4/3}$ implying $n\to 4/3$. These two asymptotic regimes define the bounds established in Eq.\eqref{eq:app:n-range}.

\begin{figure}
\centering
\includegraphics[width=8.6cm]{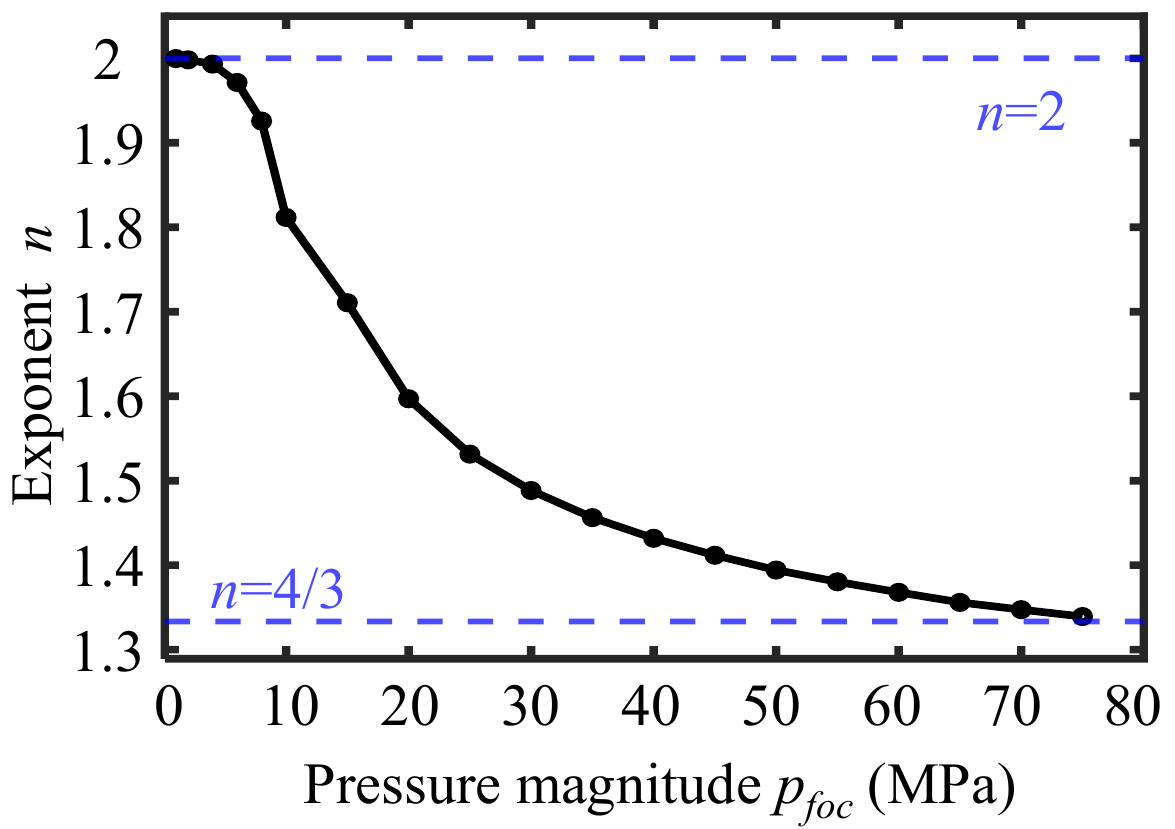}
\caption{Scaling exponent versus focal pressure.At low pressure magnitude the exponent $n$ approaches 2, at high pressure magnitude it tends towards 4/3, with a monotonic decrease across the transitional regime.}
\label{Fig:n versus p}
\end{figure}

To evaluate the pressure dependence of the local exponent $n$, we apply a centred finite difference to equation~\ref{eq:app:def-n}, using the streaming velocities obtained from finite-element (FE) simulations together with their corresponding focal pressure magnitudes. The computation is performed at a frequency of 40~MHz, with a focal length $h$=1~mm, and a transducer comprising $N$=26 turns. The resulting curve of $n(p_{\rm foc})$ is shown in~Fig.\ref{Fig:n versus p}. As expected, the exponent approaches the limiting values of 2 and 4/3 at low and high pressures, respectively. In the intermediate regime, $n$ varies continuously between these bounds, and no single, universal power law applies. These results provide strong quantitative support for the validity of the proposed scaling theory.

\bibliography{main}

@article{li2019axial,
  title={Axial acoustic field barrier for fluidic particle manipulation},
  author={Li, Nan and Kale, Akshay and Stevenson, Adrian C},
  journal={Applied Physics Letters},
  volume={114},
  number={1},
  pages  = {013702},
  year={2019},
  publisher={AIP Publishing}
}

@article{mace2008modelling,
  title={Modelling wave propagation in two-dimensional structures using finite element analysis},
  author={Mace, Brian R and Manconi, Elisabetta},
  journal={Journal of Sound and Vibration},
  volume={318},
  number={4-5},
  pages={884--902},
  year={2008},
  publisher={Elsevier}
}

@article{friend2011microscale,
  title={Microscale acoustofluidics: Microfluidics driven via acoustics and ultrasonics},
  author={Friend, James and Yeo, Leslie Y},
  journal={Reviews of Modern Physics},
  volume={83},
  number={2},
  pages={647--704},
  year={2011},
  publisher={APS}
}

@article{stokes1851effect,
  author  = {Stokes, George G.},
  title   = {On the Effect of the Internal Friction of Fluids on the Motion of Pendulums},
  journal = {Transactions of the Cambridge Philosophical Society},
  volume  = {9},
  pages   = {8--106},
  year    = {1851}
}

@article{dubrovski2023theory,
  title={Theory of acoustic streaming for arbitrary Reynolds number flow},
  author={Dubrovski, Oles and Friend, James and Manor, Ofer},
  journal={Journal of Fluid Mechanics},
  volume={975},
  pages={A4},
  year={2023},
  publisher={Cambridge University Press}
}

@article{daru2021acoustically,
  title={Acoustically induced thermal effects on Rayleigh streaming},
  author={Daru, Virginie and Weisman, Catherine and Baltean-Carl{\`e}s, Diana and Bailliet, H{\'e}l{\`e}ne},
  journal={Journal of Fluid Mechanics},
  volume={911},
  pages={A7},
  year={2021},
  publisher={Cambridge University Press}
}

@article{li2024eckart,
  title={Eckart streaming with nonlinear high-order harmonics: An example at gigahertz},
  author={Li, Shiyu and Cui, Weiwei and Baasch, Thierry and Wang, Bin and Gong, Zhixiong},
  journal={Physical Review Fluids},
  volume={9},
  number={8},
  pages={084201},
  year={2024},
  publisher={APS}
}

@article{moudjed2014scaling,
  title={Scaling and dimensional analysis of acoustic streaming jets},
  author={Moudjed, Brahim and Botton, Val{\'e}ry and Henry, Daniel and Ben Hadid, Hamda and Garandet, J-P},
  journal={Physics of Fluids},
  volume={26},
  number={9},
  pages={093602},
  year={2014},
  publisher={AIP Publishing}
}

@article{li2025reversing,
  title={Reversing the acoustic contrast factor by tuning the medium can make focused beams trap cells in three dimensions},
  author={Li, Shiyu and Gong, Zhixiong},
  journal={Physics of Fluids},
  volume={37},
  number={1},
  pages = {012003},
  year={2025},
  publisher={AIP Publishing}
}

@article{gong2022single,
  title={Single beam acoustical tweezers based on focused beams: a numerical analysis of two-dimensional and three-dimensional trapping capabilities},
  author={Gong, Zhixiong and Baudoin, Michael},
  journal={Physical Review Applied},
  volume={18},
  number={4},
  pages={044033},
  year={2022},
  publisher={APS}
}

@article{daru2024high,
  title={High-speed and acceleration micrometric jets induced by GHz streaming: A numerical study with direct numerical simulations},
  author={Daru, Virginie and Vincent, Bjarne and Baudoin, Michael},
  journal={The Journal of the Acoustical Society of America},
  volume={155},
  number={4},
  pages={2470--2481},
  year={2024},
  publisher={AIP Publishing}
}

@article{qiu2021fast,
  title={Fast microscale acoustic streaming driven by a temperature-gradient-induced nondissipative acoustic body force},
  author={Qiu, Wei and Joergensen, Jonas Helboe and Corato, Enrico and Bruus, Henrik and Augustsson, Per},
  journal={Physical review letters},
  volume={127},
  number={6},
  pages={064501},
  year={2021},
  publisher={APS}
}

@article{slama2019characterization,
  title={Characterization of focused-ultrasound-induced acoustic streaming},
  author={Slama, R Ben Haj and Gilles, Bruno and Chiekh, M Ben and Bera, Jean-Christophe},
  journal={Experimental Thermal and Fluid Science},
  volume={101},
  pages={37--47},
  year={2019},
  publisher={Elsevier}
}

@article{baudoin2020acoustic,
  title={Acoustic tweezers for particle and fluid micromanipulation},
  author={Baudoin, Micha{\"e}l and Thomas, J-L},
  journal={Annual Review of Fluid Mechanics},
  volume={52},
  pages={205--234},
  year={2020},
  publisher={Annual Reviews}
}

@article{zhou2025acoustic,
  title={An acoustic squeezer for assessment of multiparameter cell mechanical properties},
  author={Zhou, Wei and Li, Yingyin and Liu, Yifan and Quan, Hao and Li, Pengqi and Li, Fei and Niu, Lili and Zheng, Hairong and Meng, Long},
  journal={Ultrasonics},
  pages={107622},
  year={2025},
  publisher={Elsevier}
}

@article{hochmuth2000micropipette,
  title={Micropipette aspiration of living cells},
  author={Hochmuth, Robert M},
  journal={Journal of biomechanics},
  volume={33},
  number={1},
  pages={15--22},
  year={2000},
  publisher={Elsevier}
}

@article{vles1933recherches,
  title={Recherches sur une d{\'e}formation m{\'e}canique des {\oe}ufs d’Oursin},
  author={Vl{\`e}s, Fr},
  journal={Arch Zool exp gin},
  volume={75},
  pages={421--463},
  year={1933}
}

@article{ashkin1986observation,
  title={Observation of a single-beam gradient force optical trap for dielectric particles},
  author={Ashkin, Arthur and Dziedzic, James M and Bjorkholm, John E and Chu, Steven},
  journal={Optics letters},
  volume={11},
  number={5},
  pages={288--290},
  year={1986},
  publisher={Optical Society of America}
}

@article{baudoin2020spatially,
  title={Spatially selective manipulation of cells with single-beam acoustical tweezers},
  author={Baudoin, Michael and Thomas, Jean-Louis and Sahely, Roudy Al and Gerbedoen, Jean-Claude and Gong, Zhixiong and Sivery, Aude and Matar, Olivier Bou and Smagin, Nikolay and Favreau, Peter and Vlandas, Alexis},
  journal={Nature communications},
  volume={11},
  number={1},
  pages={4244},
  year={2020},
  publisher={Nature Publishing Group UK London}
}

@article{blazquez2019optical,
  title={Optical tweezers: Phototoxicity and thermal stress in cells and biomolecules},
  author={Bl{\'a}zquez-Castro, Alfonso},
  journal={Micromachines},
  volume={10},
  number={8},
  pages={507},
  year={2019},
  publisher={MDPI}
}

@article{de2012recent,
  title={Recent advances in magnetic tweezers},
  author={De Vlaminck, Iwijn and Dekker, Cees},
  journal={Annual review of biophysics},
  volume={41},
  number={1},
  pages={453--472},
  year={2012},
  publisher={Annual Reviews}
}

@article{gong2021three,
  title={Three-dimensional trapping and dynamic axial manipulation with frequency-tuned spiraling acoustical tweezers: a theoretical study},
  author={Gong, Zhixiong and Baudoin, Michael},
  journal={Physical Review Applied},
  volume={16},
  number={2},
  pages={024034},
  year={2021},
  publisher={APS}
}

@article{muller2012numerical,
  title={A numerical study of microparticle acoustophoresis driven by acoustic radiation forces and streaming-induced drag forces},
  author={Muller, Peter Barkholt and Barnkob, Rune and Jensen, Mads Jakob Herring and Bruus, Henrik},
  journal={Lab on a Chip},
  volume={12},
  number={22},
  pages={4617--4627},
  year={2012},
  publisher={Royal Society of Chemistry}
}

@article{sapozhnikov2013radiation,
  title={Radiation force of an arbitrary acoustic beam on an elastic sphere in a fluid},
  author={Sapozhnikov, Oleg A and Bailey, Michael R},
  journal={The Journal of the Acoustical Society of America},
  volume={133},
  number={2},
  pages={661--676},
  year={2013},
  publisher={AIP Publishing}
}

@article{augustsson2016iso,
  title={Iso-acoustic focusing of cells for size-insensitive acousto-mechanical phenotyping},
  author={Augustsson, Per and Karlsen, Jonas T and Su, Hao-Wei and Bruus, Henrik and Voldman, Joel},
  journal={Nature communications},
  volume={7},
  number={1},
  pages={11556},
  year={2016},
  publisher={Nature Publishing Group UK London}
}

@article{karlsen2016acoustic,
  title={Acoustic force density acting on inhomogeneous fluids in acoustic fields},
  author={Karlsen, Jonas T and Augustsson, Per and Bruus, Henrik},
  journal={Physical review letters},
  volume={117},
  number={11},
  pages={114504},
  year={2016},
  publisher={APS}
}

@article{neuman2008single,
  title={Single-molecule force spectroscopy: optical tweezers, magnetic tweezers and atomic force microscopy},
  author={Neuman, Keir C and Nagy, Attila},
  journal={Nature methods},
  volume={5},
  number={6},
  pages={491--505},
  year={2008},
  publisher={Nature Publishing Group}
}

@article{baresch2016observation,
  title={Observation of a single-beam gradient force acoustical trap for elastic particles: acoustical tweezers},
  author={Baresch, Diego and Thomas, Jean-Louis and Marchiano, R{\'e}gis},
  journal={Physical review letters},
  volume={116},
  number={2},
  pages={024301},
  year={2016},
  publisher={APS}
}

@article{yang2023acoustic,
  title={Acoustic tweezers for high-throughput single-cell analysis},
  author={Yang, Shujie and Rufo, Joseph and Zhong, Ruoyu and Rich, Joseph and Wang, Zeyu and Lee, Luke P and Huang, Tony Jun},
  journal={Nature protocols},
  volume={18},
  number={8},
  pages={2441--2458},
  year={2023},
  publisher={Nature Publishing Group UK London}
}

@article{barrow2018natural,
  title={Natural killer cells control tumor growth by sensing a growth factor},
  author={Barrow, Alexander D and Edeling, Melissa A and Trifonov, Vladimir and Luo, Jingqin and Goyal, Piyush and Bohl, Benjamin and Bando, Jennifer K and Kim, Albert H and Walker, John and Andahazy, Mary and others},
  journal={Cell},
  volume={172},
  number={3},
  pages={534--548},
  year={2018},
  publisher={Elsevier}
}

@article{landsberg2012melanomas,
  title={Melanomas resist T-cell therapy through inflammation-induced reversible dedifferentiation},
  author={Landsberg, Jennifer and Kohlmeyer, Judith and Renn, Marcel and Bald, Tobias and Rogava, Meri and Cron, Mira and Fatho, Martina and Lennerz, Volker and W{\"o}lfel, Thomas and H{\"o}lzel, Michael and others},
  journal={Nature},
  volume={490},
  number={7420},
  pages={412--416},
  year={2012},
  publisher={Nature Publishing Group UK London}
}

@article{baudoin2019folding,
  title={Folding a focalized acoustical vortex on a flat holographic transducer: Miniaturized selective acoustical tweezers},
  author={Baudoin, Michael and Gerbedoen, Jean-Claude and Riaud, Antoine and Matar, Olivier Bou and Smagin, Nikolay and Thomas, Jean-Louis},
  journal={Science advances},
  volume={5},
  number={4},
  pages={eaav1967},
  year={2019},
  publisher={American Association for the Advancement of Science}
}

@article{schiller1933uber,
  title={Uber die grundlegenden Berechnungen bei der Schwerkraftaufbereitung},
  author={Schiller, Von L},
  journal={Z. Vereines Deutscher Inge.},
  volume={77},
  pages={318--321},
  year={1933}
}

@article{bach2018theory,
  title={Theory of pressure acoustics with viscous boundary layers and streaming in curved elastic cavities},
  author={Bach, Jacob S and Bruus, Henrik},
  journal={The Journal of the Acoustical Society of America},
  volume={144},
  number={2},
  pages={766--784},
  year={2018},
  publisher={AIP Publishing}
}

@article{bruus2012acoustofluidics,
  title={Acoustofluidics 7: The acoustic radiation force on small particles},
  author={Bruus, Henrik},
  journal={Lab on a Chip},
  volume={12},
  number={6},
  pages={1014--1021},
  year={2012},
  publisher={Royal Society of Chemistry}
}

@article{muller2012acoustic,
  title={Acoustic force trapping of particles in microfluidic systems},
  author={Muller, Peter B. and Barnkob, Rune and Jensen, Mads R. and Bruus, Henrik},
  journal={Lab on a Chip},
  volume={12},
  number={22},
  pages={4617--4627},
  year={2012},
  publisher={Royal Society of Chemistry}
}

@article{sen2015mechanical_index,
  title = {Mechanical Index and Ultrasound Safety},
  author = {Şen, Sadık},
  journal = {Journal of Ultrasound in Medicine},
  volume = {34},
  number = {1},
  pages = {1--3},
  year = {2015},
  doi = {10.7863/ultra.34.1.1},
  publisher = {American Institute of Ultrasound in Medicine}
}

@article{aium2021biological_effects,
  title = {AIUM Statement on the Biological Effects of Ultrasound},
  author = {{American Institute of Ultrasound in Medicine}},
  journal = {Journal of Ultrasound in Medicine},
  volume = {40},
  number = {4},
  pages = {E1--E13},
  year = {2021},
  doi = {10.1002/jum.15742}
}

@article{li2024combined,
  title={Combined effect of acoustic radiation force and acoustic streaming for focused beams to trap cells in three dimensions},
  author={Li, Shiyu and Gong, Zhixiong},
  journal={The Journal of the Acoustical Society of America},
  volume={156},
  number={4\_Supplement},
  pages={A43--A43},
  year={2024},
  publisher={Acoustical Society of America}
}

@article{nyborg1958acoustic,
  title={Acoustic streaming near a boundary},
  author={Nyborg, Wesley L},
  journal={The Journal of the Acoustical Society of America},
  volume={30},
  number={4},
  pages={329--339},
  year={1958},
  publisher={Acoustical Society of America}
}

@incollection{nyborg1965acoustic,
  title={Acoustic streaming},
  author={Nyborg, Wesley Le Mars},
  booktitle={Physical acoustics},
  volume={2},
  pages={265--331},
  year={1965},
  publisher={Elsevier}
}

@article{nama2015numerical,
  title={Numerical study of acoustophoretic motion of particles in a PDMS microchannel driven by surface acoustic waves},
  author={Nama, Nitesh and Barnkob, Rune and Mao, Zhangming and K{\"a}hler, Christian J and Costanzo, Francesco and Huang, Tony Jun},
  journal={Lab on a Chip},
  volume={15},
  number={12},
  pages={2700--2709},
  year={2015},
  publisher={Royal Society of Chemistry}
}

@article{das2025acoustothermal,
  title={Acoustothermal effect: mechanism and quantification of the heat source},
  author={Das, Pradipta Kr and Bhethanabotla, Venkat R},
  journal={Journal of Fluid Mechanics},
  volume={1012},
  pages={A11},
  year={2025},
  publisher={Cambridge University Press}
}

@article{cicek2017acoustophoretic,
  title={Acoustophoretic separation of airborne millimeter-size particles by a Fresnel lens},
  author={Cicek, Ahmet and Korozlu, Nurettin and Adem Kaya, Olgun and Ulug, Bulent},
  journal={Scientific reports},
  volume={7},
  number={1},
  pages={43374},
  year={2017},
  publisher={Nature Publishing Group UK London}
}

@article{hasegawa2001frequency,
  title={Frequency dependence of the acoustic radiation pressure on a solid sphere in water},
  author={Hasegawa, Takahi and Kido, Tohru and Min, Chen Wei and Iizuka, Takeshi and Matsuoka, Chihiro},
  journal={Acoustical Science and Technology},
  volume={22},
  number={4},
  pages={273--282},
  year={2001},
  publisher={ACOUSTICAL SOCIETY OF JAPAN}
}

@article{zhang2011geometrical,
  title={Geometrical interpretation of negative radiation forces of acoustical Bessel beams on spheres},
  author={Zhang, Likun and Marston, Philip L},
  journal={Physical Review E—Statistical, Nonlinear, and Soft Matter Physics},
  volume={84},
  number={3},
  pages={035601},
  year={2011},
  publisher={APS}
}

@inproceedings{marston2013viscous,
  title={Viscous contributions to low-frequency scattering, power absorption, radiation force, and radiation torque for spheres in acoustic beams},
  author={Marston, Philip L},
  booktitle={Proceedings of Meetings on Acoustics},
  volume={19},
  number={1},
  pages={045005},
  year={2013},
  organization={Acoustical Society of America}
}

@article{marston2016unphysical,
  title={Unphysical consequences of negative absorbed power in linear passive scattering: Implications for radiation force and torque},
  author={Marston, Philip L and Zhang, Likun},
  journal={The Journal of the Acoustical Society of America},
  volume={139},
  number={6},
  pages={3139--3144},
  year={2016},
  publisher={AIP Publishing}
}

@article{marston2017relationship,
  title={Relationship of scattering phase shifts to special radiation force conditions for spheres in axisymmetric wave-fields},
  author={Marston, Philip L and Zhang, Likun},
  journal={The Journal of the Acoustical Society of America},
  volume={141},
  number={5},
  pages={3042--3049},
  year={2017},
  publisher={AIP Publishing}
}

@article{winckelmann2023acoustic,
  title={Acoustic radiation force on a spherical thermoviscous particle in a thermoviscous fluid including scattering and microstreaming},
  author={Winckelmann, Bj{\o}rn G and Bruus, Henrik},
  journal={Physical Review E},
  volume={107},
  number={6},
  pages={065103},
  year={2023},
  publisher={APS}
}

\end{document}